\documentclass[10pt]{article}

\usepackage[T1]{fontenc}
\usepackage[utf8]{inputenc}
\usepackage[
  a4paper,
  top=1.8cm,
  bottom=2cm,
  left=1.7cm,
  right=1.7cm,
  columnsep=0.7cm
]{geometry}

\usepackage{microtype}
\usepackage{cite}
\usepackage{graphicx}
\usepackage[table]{xcolor}
\usepackage{colortbl}

\usepackage{tabularx}
\usepackage{multirow}
\usepackage{booktabs}
\usepackage{setspace}
\usepackage{listings}
\usepackage{makecell}
\usepackage{array}

\newcolumntype{Y}{>{\raggedright\arraybackslash}X}

\usepackage{soul}
\usepackage{pifont}
\usepackage{fontawesome5}

\definecolor{backcolour}{rgb}{0.97,0.97,0.97}
\definecolor{codegreen}{rgb}{0,0.6,0}
\definecolor{codegray}{rgb}{0.5,0.5,0.5}
\definecolor{codepurple}{rgb}{0.58,0,0.82}
\definecolor{keywordblue}{rgb}{0.26,0.32,0.65}
\definecolor{rulegray}{rgb}{0.7,0.7,0.7}
\definecolor{stringred}{rgb}{0.8,0.2,0.2}

\definecolor{codeerrorbg}{rgb}{1.0,0.85,0.85}
\definecolor{codeerrorfg}{rgb}{0.6,0.0,0.0}

\newcommand{\codeerror}[1]{%
  \fboxsep=0pt
  \colorbox{codeerrorbg}{%
    \textcolor{codeerrorfg}{\textbf{\ttfamily #1}}%
  }%
}

\lstdefinestyle{mystyle}{
  backgroundcolor=\color{backcolour},
  basicstyle=\ttfamily\fontsize{6}{6}\selectfont,
  keywordstyle=\bfseries\color{keywordblue},
  commentstyle=\itshape\color{codegreen},
  stringstyle=\color{stringred},
  numberstyle=\tiny\color{codegray},
  numbers=left,
  numbersep=-2.5pt,
  captionpos=b,
  frame=single,
  rulecolor=\color{rulegray},
  breaklines=true,
  breakatwhitespace=false,
  showstringspaces=false,
  showtabs=false,
  tabsize=1,
  keepspaces=true,
  aboveskip=6pt,
  belowskip=6pt
}

\usepackage{tikz}
\usepackage[most]{tcolorbox}

\newcommand{\rectnumber}[1]{%
  \vspace{1pt}
  \noindent
  \tikz[baseline=(myanchor.base)]
    \node[
      rounded corners=2pt,
      draw=black,
      fill=blue!10,
      inner sep=2pt,
      minimum size=10pt
    ]
    (myanchor) {\bfseries\small #1};%
  \hspace{0.4em}%
}

\definecolor{lightyellow}{RGB}{255,255,240}
\definecolor{darkred}{RGB}{178,34,34}
\definecolor{deepteal}{RGB}{0,77,77}

\newcounter{remarkcounter}
\newcounter{finding}

\newtcolorbox{findingbox}[1][]{
  enhanced,
  breakable,
  colback=lightyellow,
  colframe=white,
  coltitle=black,
  fonttitle=\bfseries\scshape,
  title={Remark~\refstepcounter{remarkcounter}\theremarkcounter
         \vspace{-6pt}},
  boxrule=0pt,
  left=6pt,
  right=6pt,
  top=6pt,
  bottom=6pt,
  sharp corners,
  borderline west={2pt}{0pt}{darkred},
  drop shadow southeast,
  before skip=10pt,
  after skip=10pt,
  #1
}

\newcommand{\subheading}[1]{%
  \vspace{1pt}%
  \noindent{\textit{\textbf{#1.}}}%
}

\newcommand{\likertbar}[6]{%
  \begin{tikzpicture}[x=5cm,baseline=(base)]
    \node (base) at (0,0.08) {};

    \pgfmathsetmacro{\xA}{#1}
    \pgfmathsetmacro{\xB}{#1 + #2}
    \pgfmathsetmacro{\xC}{#1 + #2 + #3}
    \pgfmathsetmacro{\xD}{#1 + #2 + #3 + #4}
    \pgfmathsetmacro{\xE}{#1 + #2 + #3 + #4 + #5}

    \def\BorderWidth{0.6pt}

    \filldraw[
      fill=red!60!white,
      draw=#6,
      line width=\BorderWidth
    ]
      (0,0) rectangle (\xA,0.26);

    \ifdim#1pt>0.05pt
      \pgfmathsetmacro{\xMidA}{\xA/2}
      \pgfmathtruncatemacro{\percA}{round(#1*100)}
      \node at (\xMidA,0.13)
        {\scriptsize\scalebox{1}{\percA\%}};
    \fi

    \filldraw[
      fill=red!30!white,
      draw=#6,
      line width=\BorderWidth
    ]
      (\xA,0) rectangle (\xB,0.26);

    \ifdim#2pt>0.05pt
      \pgfmathsetmacro{\xMidB}{(\xA+\xB)/2}
      \pgfmathtruncatemacro{\percB}{round(#2*100)}
      \node at (\xMidB,0.13)
        {\scriptsize\scalebox{1}{\percB\%}};
    \fi

    \filldraw[
      fill=orange!10,
      draw=#6,
      line width=\BorderWidth
    ]
      (\xB,0) rectangle (\xC,0.26);

    \ifdim#3pt>0.05pt
      \pgfmathsetmacro{\xMidC}{(\xB+\xC)/2}
      \pgfmathtruncatemacro{\percC}{round(#3*100)}
      \node at (\xMidC,0.13)
        {\scriptsize\scalebox{1}{\percC\%}};
    \fi

    \filldraw[
      fill=teal!50!white,
      draw=#6,
      line width=\BorderWidth
    ]
      (\xC,0) rectangle (\xD,0.26);

    \ifdim#4pt>0.05pt
      \pgfmathsetmacro{\xMidD}{(\xC+\xD)/2}
      \pgfmathtruncatemacro{\percD}{round(#4*100)}
      \node at (\xMidD,0.13)
        {\scriptsize\scalebox{1}{\percD\%}};
    \fi

    \filldraw[
      fill=teal!80!white,
      draw=#6,
      line width=\BorderWidth
    ]
      (\xD,0) rectangle (\xE,0.26);

    \ifdim#5pt>0.05pt
      \pgfmathsetmacro{\xMidE}{(\xD+\xE)/2}
      \pgfmathtruncatemacro{\percE}{round(#5*100)}
      \node at (\xMidE,0.13)
        {\scriptsize\scalebox{1}{\percE\%}};
    \fi
  \end{tikzpicture}%
}

\newcommand{\accbar}[4]{%
  \begin{tikzpicture}[x=5.5cm,baseline=(base)]
    \node (base) at (0,0.08) {};

    \pgfmathsetmacro{\xD}{#1}
    \pgfmathsetmacro{\xC}{#1 + #2}
    \pgfmathsetmacro{\xB}{#1 + #2 + #3}
    \pgfmathsetmacro{\xA}{#1 + #2 + #3 + #4}

    \def\BorderWidth{0.6pt}

    \filldraw[
      fill=cyan!60!white,
      draw=black,
      line width=\BorderWidth
    ]
      (0,0) rectangle (\xD,0.26);

    \ifdim#1pt>0.05pt
      \pgfmathsetmacro{\xMidD}{\xD/2}
      \pgfmathtruncatemacro{\percD}{round(#1*100)}
      \node at (\xMidD,0.13) {\scriptsize\percD\%};
    \fi

    \filldraw[
      fill=cyan!40!white,
      draw=black,
      line width=\BorderWidth
    ]
      (\xD,0) rectangle (\xC,0.26);

    \ifdim#2pt>0.05pt
      \pgfmathsetmacro{\xMidC}{(\xD+\xC)/2}
      \pgfmathtruncatemacro{\percC}{round(#2*100)}
      \node at (\xMidC,0.13) {\scriptsize\percC\%};
    \fi

    \filldraw[
      fill=gray!40!white,
      draw=black,
      line width=\BorderWidth
    ]
      (\xC,0) rectangle (\xB,0.26);

    \ifdim#3pt>0.05pt
      \pgfmathsetmacro{\xMidB}{(\xC+\xB)/2}
      \pgfmathtruncatemacro{\percB}{round(#3*100)}
      \node at (\xMidB,0.13) {\scriptsize\percB\%};
    \fi

    \filldraw[
      fill=gray!70!white,
      draw=black,
      line width=\BorderWidth
    ]
      (\xB,0) rectangle (\xA,0.26);

    \ifdim#4pt>0.05pt
      \pgfmathsetmacro{\xMidA}{(\xB+\xA)/2}
      \pgfmathtruncatemacro{\percA}{round(#4*100)}
      \node at (\xMidA,0.13) {\scriptsize\percA\%};
    \fi
  \end{tikzpicture}%
}

\newcommand{\fixbar}[3]{%
  \begin{tikzpicture}[x=5cm,baseline=(base)]
    \node (base) at (0,0.08) {};

    \pgfmathsetmacro{\xA}{#1}
    \pgfmathsetmacro{\xB}{#1 + #2}
    \pgfmathsetmacro{\xC}{#1 + #2 + #3}

    \def\BorderWidth{0.6pt}

    \filldraw[
      fill=cyan!30!white,
      draw=black,
      line width=\BorderWidth
    ]
      (0,0) rectangle (\xA,0.26);

    \ifdim#1pt>0.05pt
      \node at (\xA/2,0.13)
        {\scriptsize
         \pgfmathparse{int(round(#1*100))}
         \pgfmathresult\%};
    \fi

    \filldraw[
      fill=cyan!60!green!60,
      draw=black,
      line width=\BorderWidth
    ]
      (\xA,0) rectangle (\xB,0.26);

    \ifdim#2pt>0.05pt
      \node at ({(\xA+\xB)/2},0.13)
        {\scriptsize
         \pgfmathparse{int(round(#2*100))}
         \pgfmathresult\%};
    \fi

    \filldraw[
      fill=cyan!40!green!70,
      draw=black,
      line width=\BorderWidth
    ]
      (\xB,0) rectangle (\xC,0.26);

    \ifdim#3pt>0.05pt
      \node at ({(\xB+\xC)/2},0.13)
        {\scriptsize
         \pgfmathparse{int(round(#3*100))}
         \pgfmathresult\%};
    \fi
  \end{tikzpicture}%
}

\usepackage{amssymb}
\usepackage{scalerel}
\usepackage{stackengine}

\newcounter{iloop}

\newcommand\openbigstar[1][0.7]{%
  \scalerel*{%
    \stackinset{c}{-.125pt}{c}{}{%
      \scalebox{#1}{\color{white}{$\bigstar$}}%
    }{%
      \color{orange}$\bigstar$%
    }%
  }{%
    \color{orange}\bigstar
  }%
}

\newcommand{\starrating}[1]{%
  \ensuremath{%
    \pgfmathtruncatemacro{\imax}{%
      ifthenelse(int(#1)==#1,#1-1,#1)%
    }%
    \pgfmathsetmacro{\xrest}{0.9*(1-#1+\imax)}%
    \setcounter{iloop}{0}%
    \loop
      \stepcounter{iloop}%
      \ifnum\value{iloop}<\the\numexpr\imax+1
        \color{orange}\bigstar
    \repeat
    \openbigstar[\xrest]%
    \setcounter{iloop}{0}%
    \loop
      \stepcounter{iloop}%
      \ifnum\value{iloop}<\the\numexpr5-\imax\relax
        \openbigstar[.9]%
    \repeat
    \,\color{black}\textnormal{\scriptsize #1}%
  }%
}

\usepackage[hidelinks]{hyperref}

\title{%
  Understanding the Impact of AI Code Assistants on Security API Usage:
  An Empirical Study
}

\author{%
  Zahra Mousavi\textsuperscript{1} \and
  Chadni Islam\textsuperscript{2} \and
  M. Ali Babar\textsuperscript{1} \and
  Alsharif Abuadbba\textsuperscript{3} \and
  Kristen Moore\textsuperscript{3}
  \\[0.8em]
  \small
  \textsuperscript{1}Centre for Research on Engineering Software Technologies (CREST) \& Adelaide University, Australia\\[-0.1em] 
  \small
  \textsuperscript{2}Edith Cowan University, Australia
  \qquad
  \small
  \textsuperscript{3}CSIRO's Data61, Australia
}

\date{}

\begin{document}

\twocolumn[
\begin{@twocolumnfalse}

\maketitle

\begin{abstract}

    AI code assistants are transforming software development, but their
    implications for software security remain a major concern, particularly
    in the context of security APIs. These APIs are critical for safeguarding
    software systems, yet their complexity often leads to incorrect use and
    serious vulnerabilities. Developing an evidence-based understanding of
    how AI assistants influence developers’ use of these APIs is therefore
    essential for informing effective mitigation strategies. While a few
    user studies have examined the broader impact of AI assistants on
    software vulnerabilities, the use of security APIs remains unexplored
    from a developer-centered perspective. This study addresses this gap by
    presenting the first empirical investigation into how AI code assistants
    affect professional developers’ use of security APIs. We conducted a
    study with 44 developers who completed security API programming tasks
    with and without GitHub Copilot assistance. Our findings show that,
    while Copilot improves functional correctness and marginally reduces
    certain insecure patterns, it does not significantly improve secure API
    usage. We also found that developers rarely raised security concerns
    when engaging with Copilot, and many did not recognize that their final
    implementations remained insecure. Finally, we offer recommendations
    for enhancing security awareness among developers and propose future
    research directions to support safer AI-assisted software development.
\end{abstract}

\noindent
\textbf{Keywords:}
Security API \(\cdot\)
AI Code Assistants \(\cdot\)
Software Security

\vspace{1em}

\end{@twocolumnfalse}
]

\section{Introduction}

Security Application Programming Interfaces (APIs) play a crucial role in modern software development by providing essential functionalities, such as encryption
and secure communication~\cite{mousavi2023detecting}. 
Developers rely heavily on these APIs to protect various types of applications against cyber threats. SSL/TLS APIs are a prominent example and are extensively integrated within a wide range of platforms, including web browsers, mobile applications, and cloud services, to ensure the confidentiality and integrity of data transmitted over networks~\cite{fahl2012eve}. 

\begin{figure*}[t!]
  \centering
  \includegraphics[width=.9\textwidth]{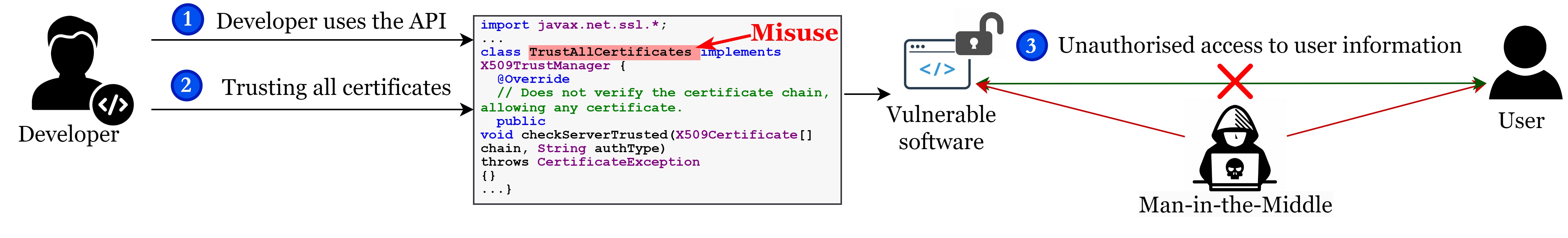}
  \caption{A misuse of an SSL/TLS API leading to the leakage of user personal information}
  \label{fig:misuse}  
\end{figure*}

However, using security APIs correctly remains a significant challenge for developers, resulting in their widespread misuse (i.e., incorrect use) across real-world software systems and open-source codebases~\cite{mousavi2023detecting,georgiev2012most,egele2013empirical,rahaman2019cryptoguard,bianchi2018broken,al2019oauthlint,kruger2019crysl}. Notably, security API misuse accounts for a substantial number of security vulnerabilities that expose systems to large-scale data breaches and significant financial losses~\cite{georgiev2012most,egele2013empirical,rahaman2019cryptoguard,bianchi2018broken,al2019oauthlint,kruger2019crysl}.
A preliminary study on non-browser software found critical misuses of SSL/TLS APIs, such as bypassing certificate validation in security-critical applications and libraries, ranging from 
payment gateways to 
mobile banking apps~\cite{georgiev2012most}. Such misuse exposes software to Man-in-the-Middle (MitM) attacks, compromising the confidentiality and integrity of network communications.
Fig.~\ref{fig:misuse} illustrates this type of misuse. A developer uses an SSL/TLS API to establish a secure connection with a server (Step \textcircled{\scriptsize 1})  but incorrectly configures it to \textit{trust all certificates} (Step \textcircled{\scriptsize 2}). This misuse enables a MitM attacker to impersonate the server, intercept the communication between a user and the application, and obtain unauthorized access to the user's information (Step \textcircled{\scriptsize 3}).

The underlying reasons for the prevalent misuse of security APIs include insufficient security training among developers~\cite{nadi2016jumping,acar2016you,acar2017developers}, inadequate or unclear API documentation~\cite{shernan2015more}, and the inherent complexity of the security APIs themselves~\cite{georgiev2012most}, which can overwhelm developers with a confusing range of configurations and options. 
These challenges often lead developers to turn to alternatives such as AI code assistants, which enable them to easily find solutions to development problems~\cite{klemmer2024using}.
Modern AI code assistants, driven by Large Language Models (LLMs), are rapidly becoming integral to developers' workflows, with many relying on them to streamline and accelerate their daily programming activities~\cite{klemmer2024using}. For example, GitHub Copilot, currently the most widely used AI code assistant, has surpassed 20 million users by mid-2025 and is used by major companies and around 90\% of Fortune 100 firms~\cite{hipolito2025github}.

This growing reliance raises serious concerns regarding the quality and security of the LLM-generated code~\cite{pearce2022asleep,siddiq2022empirical,khoury2023secure,asare2023github,sandoval2023lost,perry2023users,nair2023generating,liu2024no,kholoosi2024qualitative,asare2024user,mousavi2024investigation,fu2025security,sajadi2025llms}. LLMs like Copilot have been trained on vast code repositories containing both secure and insecure examples, posing the risk of reproducing insecure patterns. These concerns 
are especially serious in the context of security APIs, where their misuse can lead to severe vulnerabilities and compromise system security~\cite{mousavi2023detecting}. 
Therefore, it is imperative to understand \textit{how these tools influence developers’ use of security APIs in practice.}

A recent study~\cite{mousavi2024investigation} revealed that approximately 70\% of security API code generated by ChatGPT exhibited misuse patterns. 
This analysis was based solely on model outputs generated from predefined prompts, without developer involvement.
This leaves three critical gaps. \textbf{First,} existing studies lack real-world evaluations that capture developers' role in interpreting, modifying, accepting, or rejecting AI-generated code---decisions that directly influence the security of the final implementation. While a few developer-centered studies have explored the broader impact of AI assistants on security vulnerabilities~\cite{perry2023users, sandoval2023lost, asare2024user}, the developers' role in using security APIs with AI assistants remains unexplored. \textbf{Second,} it remains unclear how AI-assisted development compares to unaided coding for security APIs, both in terms of security outcomes and functional correctness, as many misuses only arise once core API calls are correctly integrated and implementations execute as intended.
\textbf{Third}, little is known about developers’ security awareness when engaging with AI assistants in security-sensitive contexts, including whether they consider security during interaction with the tool and whether they recognize security issues in the resulting code.


Therefore, conducting developer-centered studies is paramount for building an evidence-based understanding of how AI code assistants influence the use of security APIs in practice. 
This study presents the first empirical investigation of its kind into developer engagement with AI code assistants for implementing security API tasks in realistic settings. We focus on Copilot in this study, as it is currently the most widely adopted code assistant~\cite{hipolito2025github, Softonic_2024,ciodive_2024}. Our study investigates three key Research Questions (RQs): 
\begin{itemize}
\item \textit{\textbf{RQ1:}} How effectively does Copilot help developers produce functionally correct code for security API tasks?
\item \textit{\textbf{RQ2:}} How does Copilot affect the secure use of security APIs, and which misuse types does it introduce or mitigate?
\item \textit{\textbf{RQ3:}} To what extent do developers demonstrate security awareness when engaging with Copilot for security API tasks?
\end{itemize}

We conducted a user study with 44 professional developers, where each completed two development tasks involving different security APIs, one with Copilot and one without. We performed quantitative and qualitative analyses of the resulting code artifacts and developers’ engagement with Copilot. This paper makes the following contributions:
\begin{itemize}
    \item We present the first developer-centered empirical study of AI-assisted programming with security APIs, comparing Copilot-assisted and unaided development in realistic task settings.
    \item We show that Copilot can improve functional correctness, especially for more complex security API tasks, but does not enable developers to produce fully secure implementations.
    \item We identify the types of security API misuses that persist, are introduced, or are partially mitigated in Copilot-assisted development.
    \item We reveal limited security-aware engagement with Copilot: only two participants explicitly considered security in their prompts, and many failed to recognize that their final implementations remained insecure, despite being informed that security would be evaluated.
    \item We further show through a post-study prompting analysis that targeted security-focused prompts can help address some misuses, although Copilot still does not reliably prevent all of them.
\end{itemize}

Together, these findings underscore the need to foster a \textbf{\textit{security-conscious}} mindset in AI-assisted development and to support developers in critically assessing the security of AI-generated code. We further discuss the implications of these findings and provide recommendations for developers, tool designers, and researchers. All study materials are available online~\cite{material}.


The remainder of this paper is organized as follows: Section~\ref{sec:related-work} covers background and related work. Section~\ref{sec:method} details the methodology. Section~\ref{sec:results} presents results, followed by analysis in Section~\ref{sec:discussion}. Threats to validity are discussed in Section~\ref{sec:threats}, and the paper concludes in Section~\ref{sec:conclusion}.

\begin{table*}[t!]
\centering
\caption{Comparison of studies on security implications of AI-assisted coding}\vspace{5pt}
\label{tab:related_work_comparison}

\scriptsize
\setlength{\tabcolsep}{4.5pt}
\renewcommand{\arraystretch}{1.15}
\rowcolors{2}{gray!8}{white}

\begin{tabularx}{\textwidth}{
>{\raggedright\arraybackslash}p{1.75cm}
>{\raggedright\arraybackslash}p{1.35cm}
>{\raggedright\arraybackslash}p{1.85cm}
>{\raggedright\arraybackslash}p{1.60cm}
>{\raggedright\arraybackslash}p{1.85cm}
>{\raggedright\arraybackslash}X
}

\rowcolor{white}
\textbf{Study}
& \textbf{AI Model}
& \textbf{Participants}
& \textbf{Sec.\ Focus}
& \textbf{Tasks}
& \textbf{Key Findings} \\
\hline

Sandoval et al.~\cite{sandoval2023lost}
& Codex
& 58 students
& MITRE Top 25 CWEs
& Linked-list task (C)
& Improved correctness; $<10\%$ increase in critical CWEs \\

Perry et al.~\cite{perry2023users}
& Codex
& 47 students/professionals
& Web vulnerabilities
& 5 security tasks (Py/JS/C)
& Less secure code; security overestimation \\

Asare et al.~\cite{asare2024user}
& Copilot
& 25 students/professionals
& 13 CWEs (e.g., SQLi)
& 2 real-world C tasks
& Improved security only for complex tasks \\

Fu et al.~\cite{fu2025security}
& Copilot, CodeWhisp., Codeium
& None (GitHub artifacts)
& 43 CWEs
& Py/JS AI snippets
& 30\% vulnerable snippets; includes 8 CWE Top 25 \\

Mousavi et al.~\cite{mousavi2024investigation}
& ChatGPT
& None (model-based)
& API misuse
& 48 Java tasks (5 APIs)
& 70\% outputs contained API misuses \\

\textbf{Our Study}
& Copilot
& 44 professionals
& API misuse
& 2 Java tasks
& Improved correctness; no significant security gain; security often overlooked and overestimated \\

\end{tabularx}

\normalsize
\end{table*}

\section{Related work}
\label{sec:related-work}
Several studies have raised significant concerns about the security implications of AI assistants for code generation~\cite{pearce2022asleep, siddiq2022empirical, khoury2023secure, asare2023github, sandoval2023lost, perry2023users, nair2023generating, liu2024no, kholoosi2024qualitative, asare2024user, mousavi2024investigation, fu2025security, sajadi2025llms}. An early study by Pearce et al.~\cite{pearce2022asleep} showed that approximately 40\% of Copilot's suggestions contained security flaws. Similar studies have confirmed that AI-generated code can inherit insecure patterns from training data~\cite{pearce2022asleep, siddiq2022empirical, nair2023generating, khoury2023secure, liu2024no, sajadi2025llms}. However, these analyses were largely conducted in controlled environments using predefined prompts without involving developers. Recognizing this gap, recent studies have shifted toward evaluations in real-world contexts.
Fu et al.~\cite{fu2025security} examined code snippets generated by developers using Copilot and two other AI tools in GitHub projects, finding that approximately 30\% contained security weaknesses spanning 43 CWE types, including eight from the CWE Top 25 list. To assess Copilot’s performance relative to human developers, Asare et al.~\cite{asare2023github} used a dataset of vulnerabilities introduced in human-written code to construct prompts. They found that Copilot reproduced the same vulnerabilities in only about 33\% of cases, suggesting that while it generates insecure code, it does not perform worse than developers.

User studies have also examined the security implications of AI assistance during coding, though findings vary by tool and task complexity. Sandoval et al.~\cite{sandoval2023lost} found that Codex improved functional correctness while increasing critical security bugs by at most 10\%. In contrast, Perry et al.~\cite{perry2023users} found that 47 students and professionals using Codex produced significantly less secure code and were more likely to overestimate its security. Focusing on Copilot, Asare et al.~\cite{asare2024user} studied 25 students and professionals and found that Copilot improved security only for more complex problems, with no significant effect on simpler ones or on specific vulnerability categories.

Notably, \textit{none of the aforementioned studies specifically addressed the critical issue of security API misuse}, which remains a major concern in secure software development. Compared with general vulnerabilities, security API misuse places greater emphasis on compliance with API-specific security constraints and often leads to severe security consequences. Mousavi et al.~\cite{mousavi2024investigation} investigated this issue using ChatGPT and found misuses in 70\% of generated code instances for security API tasks. However, their study relied on researcher-crafted prompts and did not involve professional developers. This is a key limitation, as developers ultimately shape the final code by accepting, modifying, or rejecting AI-generated suggestions~\cite{arani2023sok, arani2024systematic}. Further, its focus on ChatGPT leaves a gap in understanding how Copilot---the most widely adopted AI assistant among developers---affects security API usage in practice. Our study addresses these gaps through a developer-centered, real-world evaluation. Specifically, \textit{(1) we involve professional developers in realistic programming tasks to assess security API usage in practice; (2) we focus on Copilot given its widespread adoption; and (3) we analyze developers’ consideration of security in their prompts and its impact on final code security.} Table~\ref{tab:related_work_comparison} outlines the key differences between our study and prior work. \textit{\textbf{To our knowledge, this is the first in-depth study of how GitHub Copilot influences security API usage in real-world development settings.}}

\begin{figure*}[t!]
    \centering
    \includegraphics[width=.9\linewidth]{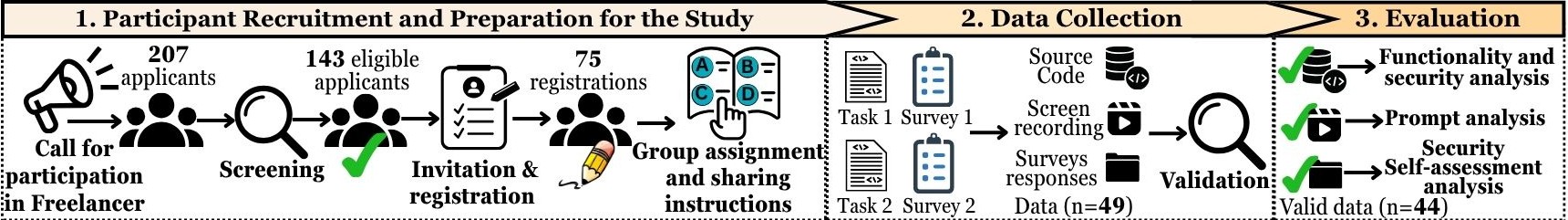}
    \caption{An overview of the research methodology for the user study}
    \label{fig:overview}
\end{figure*}

\section{Study Design}\label{sec:method} 
This section outlines the design of our empirical study, including the programming tasks, participant recruitment process, study procedure, experimental setup, and evaluation methodology. An overview of the research methodology is illustrated in Fig.~\ref{fig:overview}. The following subsections elaborate on each component, along with a discussion of the ethical considerations relevant to the study. 
\vspace{-10pt}

\subsection{Task Design}
To design effective programming tasks for our study, we aimed to achieve three primary goals. \textbf{First}, tasks were crafted to reflect real-world security challenges that developers commonly face in practice. \textbf{Second}, they were specifically designed to expose developers to common security API misuses, allowing us to observe whether participants would address them. \textbf{Finally}, we ensured that tasks are feasible to complete within a reasonable timeframe. We selected Java as the target language due to its widespread adoption in software development and the inherent complexity of its security APIs, which are frequently prone to misuse by developers in practice~\cite{java_popularity, PYPL}.

To fit within the practical time limits in a controlled user study while enabling in-depth analysis, we limited our scope to two security APIs that represent distinct security functionalities and allow observation of diverse misuse patterns across realistic scenarios. We focused on two widely used and security-critical Java APIs: 
\textbf{\textit{(i) Java Secure Socket Extension (JSSE)}}, which enables secure communication over SSL/TLS protocols~\cite{fahl2012eve, georgiev2012most, nair2023generating}, and 
\textbf{\textit{(ii) Google OAuth}}, which enables applications to access Google user data on their behalf without exposing their credentials~\cite{hardt2012rfc, al2019oauthlint}. 
While this selection does not cover the full spectrum of security APIs, it reflects two major classes of security functionality, secure communication and delegated authorization, both of which are commonly misused and associated with high-impact vulnerabilities~\cite{mousavi2023detecting}.

Each of these APIs formed the basis of one of the two programming tasks assigned in our study.
The first task required participants to use JSSE to implement a \texttt{\small{createSSLSocket}} method. This method needs to configure an SSL/TLS socket that allows a client application to securely connect to a server. 
This task represents one of the most frequent patterns for secure communication and exposes developers to well-known misuses such as improper protocol selection, certificate validation, and hostname verification~\cite{fahl2012eve, georgiev2012most, nair2023generating}.
The second task introduced a more complex programming scenario. Participants were required to implement an \texttt{\small{authorize}} method using the Google OAuth API as part of a desktop application. The application was designed to retrieve the number of unread messages in a user's Gmail inbox, subject to user consent. 
This task captures a realistic OAuth integration workflow and exposes participants to frequent authorization-flow misuses identified in prior work~\cite{hardt2012rfc, al2019oauthlint}.
For both tasks, we provided participants with a skeleton application that included stub code for non-security-related functionalities. Participants were then asked to implement the essential security components. Full task descriptions are available in the online supplementary material~\cite{material}.

\subsection{Recruitment and Participant Pool}
Following prior security-focused work~\cite{asare2024user}, we recruited professional developers aged 18+ with $\geq$1 year of Java experience. Prior security knowledge was not required as we aimed to examine how Copilot influences developers’ perceptions and experiences with security APIs regardless of their background. In addition, we required familiarity with VS Code, the IDE used in our study.
We selected Freelancer as our recruitment platform, based on recommendations from prior research highlighting its effectiveness in recruiting participants for security-focused development studies~\cite{kaur2022recruit}. Freelancer has also been used in several user studies involving security programming tasks~\cite{naiakshina2020conducting, naiakshina2019if, geierhaas2022let, danilova2020replication}.

Over a period of nearly three months, we posted a project on Freelancer in multiple iterations. The project description clearly outlined the eligibility criteria and provided a concise overview of the study. While we mentioned that the study involved security programming tasks, we did not disclose the specific APIs under evaluation. To expand our reach, we also used Freelancer’s recruiter service, which directly contacts freelancers with relevant skill sets and invites them to apply. Additionally, we encouraged participants to share the project with other eligible freelancers within their networks.

Interested freelancers submitted bids with a payment offer and a short proposal. In total, we received 207 applications. A multi-stage screening process was implemented to select eligible candidates. First, we excluded applicants with a rating below 4.5 (unless new to the platform), and those requesting higher compensation than offered. Next, we reviewed the proposals, excluding those that were irrelevant to the project. We then verified that applicants met the eligibility requirement of at least one year of Java experience. If a developer’s profile did not clearly demonstrate Java expertise, we requested additional supporting materials, such as links to GitHub repositories showcasing their Java projects.

\begin{figure*}[!t]
    \centering
    \includegraphics[width=.95\linewidth]{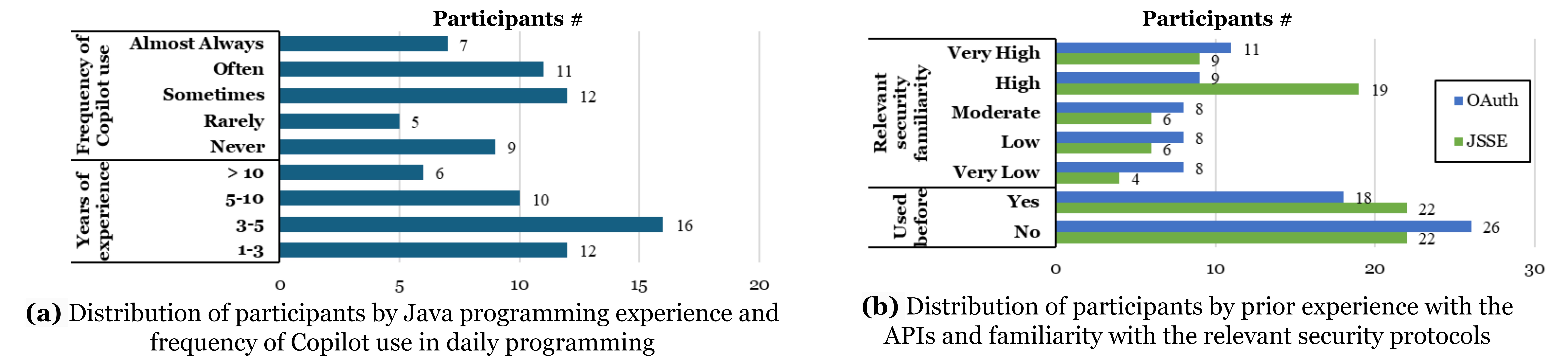}
    \caption{
    Participant overview (n=44): (a) Java experience and Copilot use frequency; (b) prior API experience and familiarity with the corresponding security protocol.}
    \label{fig:demo}
\end{figure*}

Following this screening process, 143 applicants were invited to participate. Each invited participant received a participant information sheet detailing the study, a consent form, and a link to a registration form that collected their demographic information and allowed them to select a convenient time for participation. As a result, 75 registered, of whom 11 later withdrew from the study, and 15 failed to attend their scheduled sessions. Of the 49 who joined the study and completed the programming tasks, five were excluded due to non-compliance with study guidelines, such as suspicious use of AI tools on the task that was intended to be completed independently without AI assistance. Ultimately, data from 44 participants were included in our final analysis.
Each participant who completed the study tasks received a compensation of AUD \$100. 
Fig.~\ref{fig:demo} shows a demographic overview of the final participants. 
Further details on anonymized participants and recruitment materials are available online~\cite{material}.

\subsection{Study Procedure}
We conducted a within-subject study~\cite{lazar2017research} in which all participants completed tasks under both experimental and control conditions. In the experimental condition, participants completed a task with Copilot assistance, whereas in the control condition, they completed a different task without any AI tool. 
To mitigate 
learning and fatigue effects, we counterbalanced Copilot usage (enabled vs. disabled) and task order (JSSE first vs. OAuth first). This led to four groups (2 conditions × 2 task orders), labeled A, B, C, and D, with 11 participants each. For example, Group A completed the JSSE task with Copilot, followed by the OAuth task without Copilot. The remaining groups followed different combinations of task order and Copilot availability, ensuring that for each task, half of the participants used Copilot while the other half did not. Group assignment followed permuted block randomization (block size = 4), and 
each participant was assigned an ID based on their group and number (e.g., A1--11, B1--11).

To replicate a realistic coding environment, participants in both settings were allowed to access internet resources. Those in the control condition were restricted from using any AI assistants, including Copilot, while those in the experimental condition had access to all Copilot features (e.g., inline completions, chat interface, and fixes).
To ensure familiarity with the tool, all participants were asked to review a brief tutorial before their sessions.

Since developers need clear prompts to write secure code~\cite{naiakshina2019if}, participants were informed that \textit{their solutions would be evaluated for both correctness and security}. They were given 2 hours to complete both tasks, with an optional 30-minute break after the first one. Although full functionality was not required for compensation, they were encouraged to complete as much as possible within this time. After each task, participants completed a short survey that asked whether this was their first experience with the API used, about their familiarity with the relevant security protocol, and whether they believed they had solved the task securely. With participants’ consent, all sessions were screen-recorded for compliance monitoring and later analysis. 

\subsection{Experimental Infrastructure}
All participants completed the study within a controlled virtual environment using Amazon Web Services (AWS) virtual machines (VMs), each configured with 16 vCPUs, 64 GiB of RAM, and running Ubuntu 20.04 LTS. We pre-installed VS Code, the required Java packages, and GitHub Copilot (
Version v1.250.0, running on GPT-4o, the latest available version at the time of conducting the study). 
To support the study design, we prepared four distinct VM configurations corresponding to the four participant groups. For each group, two separate VS Code instances---Copilot enabled and disabled---were provided
to ensure Copilot usage was aligned with the assigned condition for each task. Each VM included a README file containing study instructions, as well as the specific task order based on the participant’s group assignment.
Participants were assigned a unique VM instance from their respective group, which they accessed remotely from their personal devices to complete the tasks during their scheduled time.

\subsection{Evaluation Methodology}
Our evaluation involved \textit{(i) Code Analysis} (\textit{RQ1} and \textit{RQ2}) and \textit{(ii) Security Awareness Analysis} (\textit{RQ3}), as detailed below.

\subsubsection{Code Analysis.}
For each task, we analyzed the functionality and security of code samples. 
Functionality was assessed by whether the solutions executed without errors and produced the expected outputs.
Solutions that required only minor adjustments to become functional were categorized as semi-functional. Only functional and semi-functional solutions proceeded to the security evaluation.
The security evaluation focused on identifying instances of security API misuse within code samples. This assessment was guided by the taxonomy of security API misuses established by Mousavi et al.~\cite{mousavi2023detecting}. Each implementation was analyzed for the presence of misuse patterns defined in this taxonomy.

Several tools exist for detecting security API misuse, but they often produce high rates of false positives and negatives, limiting their applicability to our context. Preliminary tests with CryptoGuard~\cite{rahaman2019cryptoguard}, a tool recognized for its precision, failed to effectively detect JSSE misuses relevant to our study. Additionally, no automated tools currently exist to detect OAuth misuses. 

Given these limitations, 
the first author and a software security researcher from our lab, each with four years of security experience, independently conducted a manual review of all participant submissions. Each reviewer assessed the implementations against our predefined misuse criteria and recorded the presence or absence of each misuse category. We observed a high level of inter-rater agreement ($\kappa$ = 0.97), indicating strong consistency between the two reviewers’ assessments. Any disagreements were subsequently resolved through discussion, with both reviewers revisiting the corresponding code and reaching a final consensus classification.

To statistically examine the effects of Copilot use and participants’ background on our results, we applied logistic regression~\cite{hosmer2013applied}. Logistic regression is well suited for modeling binary dependent outcomes (e.g., correct vs. incorrect) and enables us to estimate the influence of multiple predictors simultaneously. In our models, the primary predictors were the Copilot condition (with vs. without Copilot) and participants’ background characteristics. We fit separate models for each task, such that each participant contributed a single observation per model. This design ensured there were no repeated measures within a given model, and therefore satisfied the independence assumption required for logistic regression.



\subsubsection{Security Awareness Analysis.}

To investigate developers’ security awareness, we analyzed (i) prompts to assess whether participants considered security during interaction, and (ii) post-task self-assessments to determine whether they recognized security problems in their final implementations.
\paragraph{Prompt Analysis.}

This analysis focused exclusively on the experimental condition in which participants used Copilot.
Specifically, we examined the natural-language prompts participants used to interact with Copilot to assess whether they expressed any security-related intentions or concerns.
Out of 44 participants, the chat session histories were available for 38. We extracted these logs directly from Copilot’s chat interface. For the remaining 6 participants, the chat histories were unavailable because the developers had closed their Copilot Chat sessions, and Copilot does not persist conversation histories across sessions. In these cases, we manually reviewed participants' screen recordings to capture their prompts.
Additionally, the participants could interact with Copilot through inline chat prompts within the code editor or by embedding natural-language instructions as comments to receive code suggestions. Since these interactions are not recorded in the chat interface, we again relied on screen recordings to manually extract such prompts.
All collected prompts were manually analyzed to identify any indications of security awareness. We also reviewed Copilot’s responses and participants’ final implementations to assess how security-oriented prompts affected the generated code, particularly regarding API misuses.
\paragraph{Self-Assessment Analysis.}

To investigate whether participants recognized security issues in their implementations, we analyzed post-task self-assessments collected after each task. 
Participants rated their agreement with the statement \textit{“I believe that I solved this task securely”} on a 5-point Likert scale ranging from \textit{Strongly Disagree} to \textit{Strongly Agree}. In this study, responses of \textit{Disagree} and \textit{Strongly Disagree} were interpreted as explicit recognition that the solution was insecure. These responses were then compared against the actual security outcomes identified through our misuse analysis.
\subsection{Pilot Study}
Prior to the main study, we conducted a pilot study with six developers, including two from our lab and four recruited via Freelancer. The pilot aimed to assess the clarity of the task instructions and the feasibility of completing the programming tasks within the allotted time. After completing the tasks, participants filled out a short survey about their experience, including any difficulties in understanding the tasks or using the study environment. We also gathered additional qualitative feedback through follow-up discussions, conducted via Freelancer chat for remote participants and in person for lab-based participants. Based on this feedback, we refined the task descriptions, added explicit completion criteria, expanded the step-by-step navigation instructions for the study environment, and addressed technical issues identified during the pilot such as screen recording interruptions. 
\subsection{Ethics Approval}
This study was approved by our organization’s Human Research Ethics Committee and conducted in full compliance with its ethical guidelines.
\section{Experimental Results}
\label{sec:results}
In this section, we present our findings for each research question.
\subsection{RQ1: Functionality Analysis}
This section addresses RQ1 by examining the extent to which Copilot helps developers produce functionally correct code. We analyzed participants’ success in completing the assigned tasks and assessed how AI assistance influenced their performance. For the simpler JSSE task, all participants produced functional code regardless of Copilot use, although those with Copilot completed the task faster on average (14 minutes vs.\ 26 minutes). In contrast, the OAuth task was more challenging, particularly without AI assistance. In the control condition, only 45\% of participants (10/22) produced functional code, with an average completion time of 62 minutes. With Copilot, the success rate rose to 91\% (20/22), while the average completion time decreased by 63\% to 23 minutes.

According to the logistic regression analysis, Copilot had a statistically significant positive effect on functional outcomes (coefficient $= 3.36$, $p = 0.007$), indicating that participants who used Copilot were significantly more likely to complete the task successfully than those who did not, even after controlling for background factors. Among the background predictors, programming experience ($p = 0.041$) and prior API experience ($p = 0.046$) were significantly associated with functional outcomes, whereas security familiarity and Copilot use frequency showed positive but statistically non-significant trends. 

\begin{figure*}
    \centering
    \includegraphics[width=.9\linewidth]{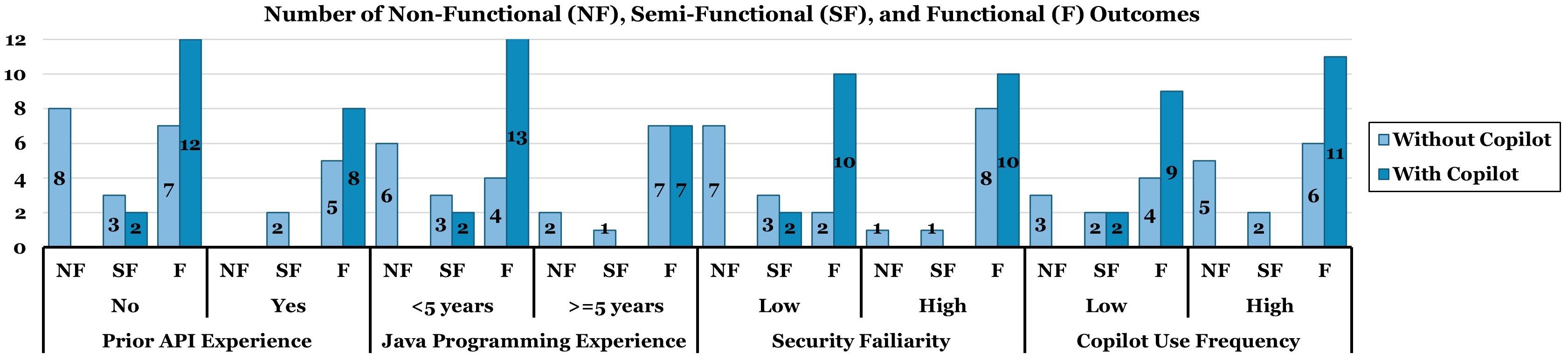}
    \caption{
    Distribution of OAuth task functionality levels by background. 
    \scriptsize 
    Copilot use freq and security familiarity were rated on a 5-point scale and grouped into low (1--3) and high (4--5).}
    \label{fig:func}
\end{figure*}

We also reviewed non-functional samples and identified semi-functional implementations for inclusion in the subsequent security analysis. Fig.~\ref{fig:func} presents the distribution of non-functional, semi-functional, and functional OAuth implementations across participant background levels.

\begin{findingbox}
\itshape\small
• All participants
completed the JSSE task with functional solutions.\\
• Copilot significantly improved functionality for the more complex OAuth task.
\end{findingbox}

\subsection{RQ2: Security Analysis}
\begin{figure*}[] 
 \centering
 \includegraphics[width=.9\linewidth]{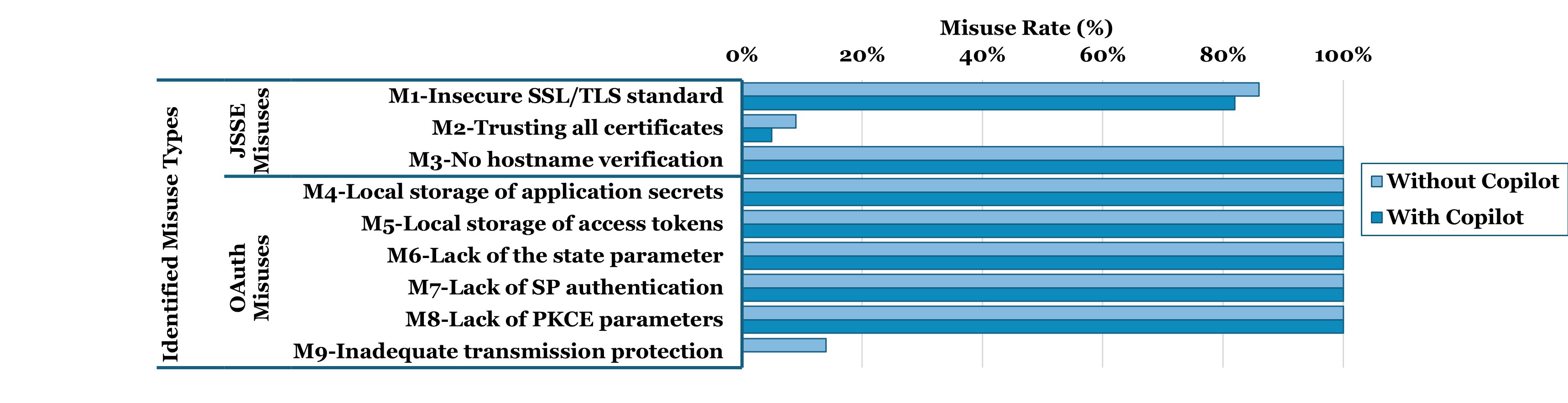}
 \caption{Misuse types and their rates across code samples; for M5, the analysis is based only on implementations that include token storage.}
 \label{fig:misuse_type}
\end{figure*}

This section addresses RQ2 by analyzing security outcomes. Our analysis revealed a concerning trend: \textit{none of the implementations were fully secure, regardless of Copilot use or participant background}. This is particularly concerning because participants had been clearly informed that their work would be evaluated for both functionality and security, and the task descriptions explicitly highlighted the security-sensitive nature of the tasks. Fig.~\ref{fig:misuse_type} presents the identified misuses and their corresponding rates. We define misuse rate as the percentage of analyzed implementations exhibiting a given misuse. We next discuss the misuses identified for each API, assigning each a unique identifier (\textit{M\#}).

\subsubsection{JSSE.}
Our study found 3 JSSE misuses as follows.\vspace{5pt}

\begin{figure*}[t!] 
    \centering
    \includegraphics[width=.9\linewidth]{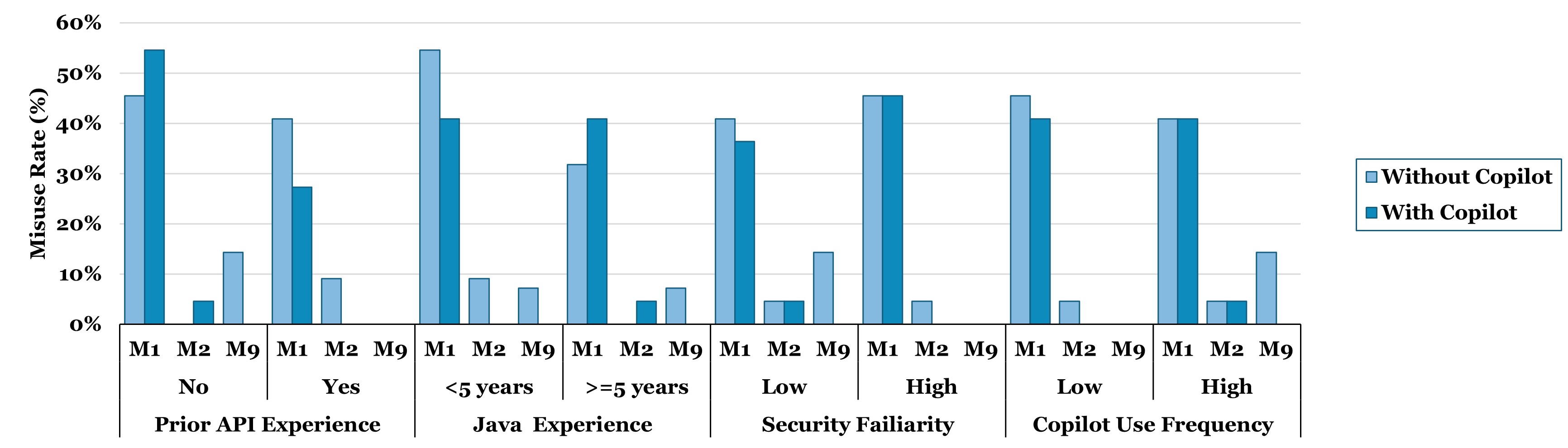}
    \caption{
    Occurrence rates of M1, M2, and M9 across backgrounds. Other misuses were observed among participants of all backgrounds, regardless of Copilot use.}
    \label{fig:misuse_bg}
\end{figure*}

\subheading{M1: Insecure SSL/TLS standard}
SSL and older TLS versions (1.0/1.1) are vulnerable to attacks such as POODLE, BEAST, and CRIME, and are therefore considered insecure~\cite{barnes2015deprecating, k2021deprecating, turner2011prohibiting}. These protocols have been deprecated, with TLS 1.2 as the minimum recommended secure version. In our study, 19 participants in the control condition and 18 in the AI-assisted condition either used outdated versions or failed to specify the TLS version, potentially allowing insecure defaults. Listing~\ref{lst:jsse_control} (line 9) shows a control-condition example using SSL, while Listing~\ref{lst:jsse_assisted} shows an AI-assisted example in which the TLS version is not specified. Fig.~\ref{fig:misuse_bg} presents the rates of this misuse across background levels, with and without Copilot. In addition to Copilot use, prior API experience showed a trend toward lower misuse rates, although the effect was not statistically significant.

\subheading{M2: Trusting all certificates}
The default JSSE trust manager validates certificates against the system trust store. In our study, we added our server's certificate to the trust store on all machines, allowing participants to rely on the default trust manager for secure validation. Most participants used this mechanism. However, two participants in the control condition and one in the AI-assisted condition implemented custom trust managers that accepted all certificates, bypassing authentication and enabling MitM attacks. Listing~\ref{lst:jsse_control} shows a developer using an empty certificate-validation method, thereby accepting all certificates as valid. Fig.~\ref{fig:misuse_bg} shows the distribution of M2 across background levels, although no meaningful trends can be drawn given the small number of cases.

\subheading{M3: Missing hostname verification}
Hostname verification is a critical security measure that ensures the hostname in the SSL certificate matches the server hostname the client is connecting to. Without it, an attacker can intercept communication by presenting a certificate for a malicious server. In our study, none of the implementations performed hostname verification. Listings~\ref{lst:jsse_control} and~\ref{lst:jsse_assisted} show examples from both conditions in which hostname verification was omitted.
\begin{lstlisting}[language=java, label={lst:jsse_control}, caption={JSSE code sample (control group) with M1 (L9), M2 (L2), and M3.}, escapeinside=$$, breaklines=true]
 private static SSLSocket createSSLSocket(String host, int port) throws UnknownHostException, IOException {
    TrustManager[] $\codeerror{trustAllCerts}$ = new TrustManager[]{new X509TrustManager() {  
    @Override
    public java.security.cert.X509Certificate[] getAcceptedIssuers() { return null; } 
    @Override
    public void checkClientTrusted(java.security.cert.X509Certificate[] certs, String authType) {} 
    @Override
    public void checkServerTrusted (java.security.cert.X509Certificate[] certs, String authType) {}  }};
    try { SSLContext sc = $\codeerror{SSLContext.getInstance("SSL");}$
          sc.init(null, trustAllCerts, new java.security.SecureRandom());
          HttpsURLConnection.setDefaultSSLSocketFactory(sc.getSocketFactory());
          SSLSocketFactory factory = (SSLSocketFactory) SSLSocketFactory.getDefault(); 
          SSLSocket s =(SSLSocket) factory.createSocket(host, port);
          s.startHandshake();
          return s;
    } catch (GeneralSecurityException e) { System.out.println(e.getStackTrace());  }
    return null; }
\end{lstlisting}
\begin{lstlisting}[language=java, label={lst:jsse_assisted}, caption={JSSE code sample (AI-assisted) with M1 (no TLS version) and M3.}, escapeinside=@@]
 private static SSLSocket createSSLSocket(String host, int port) {
    try {
    // Create an SSL context
    SSLContext sslContext =SSLContext.getInstance("TLS");
    // Get the default trust manager
    TrustManagerFactory trustManagerFactory = TrustManagerFactory.getInstance(TrustManagerFactory.getDefaultAlgorithm());
    trustManagerFactory.init((java.security.KeyStore)null);
    TrustManager[] trustManagers = trustManagerFactory.getTrustManagers();
    // Initialize with the default trust manager
    sslContext.init(null, trustManagers, new java.security.SecureRandom());
    // Create and return the SSL socket
    return (SSLSocket) sslContext.getSocketFactory().createSocket(host, port);
    } catch (NoSuchAlgorithmException | KeyManagementException | KeyStoreException | IOException e) {
        e.printStackTrace();
        return null; }}
\end{lstlisting}

\subsubsection{Google OAuth.}
Our study identified 6 critical misuses of OAuth, with all occurring in the control condition and 5 in the AI-assisted implementations. \\

\subheading{M4: Local storage of application secrets} 
Application secrets used to authenticate with the Service Provider (SP), Google in this case, must be stored securely. In our study, participants either hardcoded secrets (Listing~\ref{lst:oauth_control}) or stored them in local resource files without encryption (Listing~\ref{lst:oauth_assisted}), enabling attackers to impersonate legitimate applications if the secrets are compromised~\cite{chen2014oauth, al2019oauthlint}.
\begin{lstlisting}[language=java, label={lst:oauth_control}, caption=\footnotesize{OAuth sample (control group) with M4-9; M4 (L2), M5 (L6), M9 (L6)}, escapeinside=$$]
 private static String CLIENT_ID = "3430 ...";
 private static String CLIENT_SECRET = $\codeerror{"GOCS ..."}$;
 private static JsonFactory JSON_FACTORY = JacksonFactory.getDefaultInstance();
 // Authorize using OAuth 2.0 with provided scopes
 private static Credential authorize(Collection<String> scopes) throws Exception {
    GoogleAuthorizationCodeFlow flow = new GoogleAuthorizationCodeFlow.Builder(new $\codeerror{NetHttpTransport()}$, JSON_FACTORY, CLIENT_ID, CLIENT_SECRET, scopes).setDataStoreFactory(new $\codeerror{FileDataStoreFactory(new File("tokens"))}$).setAccessType("offline").setApprovalPrompt("force").build();
    LocalServerReceiver receiver = new LocalServerReceiver.Builder().setPort(8080).build();
    Credential credential = new AuthorizationCodeInstalledApp(flow, receiver).authorize("user");
    return credential; }
\end{lstlisting}
\begin{lstlisting}[language=java, label={lst:oauth_assisted}, caption=\vspace{-1pt}\footnotesize{OAuth sample from the assisted condition with M4--8; M4 (L4), M5 (L8)}, escapeinside=$$]
 private static Credential authorize(String[] scopes) throws Exception {
    JsonFactory JSON_FACTORY = JacksonFactory.getDefaultInstance();
    // Load client secrets.
    InputStream in = Main.class.getResourceAsStream( $\codeerror{"/client\_secrets.json"}$); 
    if (in == null) throw new FileNotFoundException("Resource not found: /client_secrets.json");
    GoogleClientSecrets clientSecrets = GoogleClientSecrets.load(JSON_FACTORY, new InputStreamReader(in));
    // Build flow and trigger user authorization request.
    GoogleAuthorizationCodeFlow flow = new GoogleAuthorizationCodeFlow.Builder(   GoogleNetHttpTransport.newTrustedTransport(), JSON_FACTORY, clientSecrets, List.of(scopes)).setDataStoreFactory(new $\codeerror{FileDataStoreFactory(new File("tokens"))}$)
   .setAccessType("offline").build();
    LocalServerReceiver receiver = new LocalServerReceiver.Builder().setPort(8888).build();
    return new AuthorizationCodeInstalledApp(flow, receiver).authorize("user"); }
\end{lstlisting}

\subheading{M5: Local storage of access tokens}
Secure storage is critical for access tokens, as they grant access to protected user resources~\cite{hardt2012rfc}. To keep the task manageable, participants were instructed to obtain, not store, tokens. Nonetheless, some implemented storage, and in all cases tokens were stored locally without encryption (Listings~\ref{lst:oauth_control} and~\ref{lst:oauth_assisted}), exposing them to unauthorized access~\cite{al2019oauthlint}. Although storage was not the focus of our evaluation, this misuse is noteworthy as it reflects common real-world practices, with no observable impact from Copilot.

\subheading{M6: Lack of the state parameter}
The \mbox{\texttt{\small state}} parameter is essential for ensuring request authenticity and protecting user sessions against CSRF attacks. OAuth guidelines recommend generating and validating a unique \mbox{\texttt{\small state}} linked to the user's session~\cite{hardt2012rfc}. However, none of the analyzed programs—whether from the control or assisted condition—implemented this security measure.

\subheading{M7: Lack of SP authentication}
OAuth transactions require mutual authentication between applications and SPs~\cite{wang2015vulnerability}. Yet, none of the analyzed programs from either group implemented SP authentication. 

\subheading{M8: Lack of PKCE parameters for authorization code grant}
OAuth security is highly impacted by the chosen grant type. All implementations 
used the authorization code grant, a generally secure type, but it remains vulnerable to code interception attacks~\cite{sharif2022best}. Current best practices 
recommend the authorization code flow with Proof Key for Code Exchange (PKCE) to ensure that the requesting application is the same one that initially requested it~\cite{sakimura2015proof}. However, no implementations in our study included PKCE. 

\subheading{M9: Inadequate transmission protection}
Ensuring secure communication throughout the OAuth process is essential for its security. However, two participants in the control condition employed HTTP without SSL protection, thereby compromising transmission security (one example shown in Listing~\ref{lst:oauth_control}). In contrast, all other participants used SSL/TLS for encrypted communication during OAuth transactions.  Fig.~\ref{fig:misuse_bg} shows the distribution of M9 across background levels, though the small number of cases limits reliable interpretation of trends.


\begin{findingbox}
\itshape\small
• No implementation was fully secure, and Copilot had \textbf{no significant impact} on misuse types.\\
• 3 JSSE misuses were found;
Copilot slightly reduced the frequency of two (M1--2).\\
• 6 OAuth misuses were found: 5 (M4--8) in both groups and 1 (M9) only in the control group ($2\times$). 
\end{findingbox}

\vspace{10pt}
\subsection{RQ3: Security Awareness Analysis}
In this section, we address RQ3 by analyzing participants’ security awareness through their prompts and self-assessments, as detailed below.

\paragraph{Prompt Analysis.}

We examined the language of participants’ prompts to assess whether they explicitly raised security-related concerns when interacting with Copilot. Although participants had been informed that their code would be evaluated for both functionality and security, only seven participants made references to security in their prompts. In five of these cases, the security-related phrasing was directly copied from the JSSE task description. For example, participant A9 prompted Copilot with: ``\textit{Give me the body of the createSSLSocket method in the code. The method needs to create and return a properly configured SSL/TLS socket to securely connect to the server}''. The second sentence, which highlights a secure connection, is a verbatim excerpt from the task description. Code outputs in response to such prompts included one or two of the three JSSE misuses targeted in our analysis.
Beyond these cases, where security-related language has been included either intentionally or unintentionally, two participants, C9 with \textit{high} and D6 with \textit{very high} self-rated security familiarity, demonstrated clear and deliberate consideration of security in their prompts. This suggests that \textit{developers with stronger security backgrounds may be more inclined to raise security concerns}, though the small number of such cases prevents drawing meaningful conclusions.

Participant C9, after successfully generating a functionally correct implementation of the OAuth task, prompted Copilot to ``\textit{review the authorize method from a security perspective}''. Copilot responded with several security recommendations like \textit{``Use secure storage mechanisms for storing credentials, such as encrypted files or secure vault services''}, which could help address the misuse related to M5, \textit{local storage of access tokens}. However, despite showing security awareness, the participant left the task at this point without addressing the existing misuses, including M5.
Following OpenAI's prompting guidelines, Participant D6 
assigned Copilot a predefined role: ``\textit{You are a tech architect and security expert. You need to implement the createSSLSocket method …}''. Despite being assigned the role of a security expert, the initial response by Copilot contained all three JSSE misuses, including M2, \textit{trusting all certificates}.
However, the response also provided a relevant security recommendation: \textit{``Make sure to replace the trust manager with a proper implementation for production use to ensure secure communication''}. Interestingly, the participant followed up with another prompt asking Copilot to ``\textit{double check and make changes if required from a security pov}''. In response, Copilot corrected M2 with a more secure setup using the default trust manager for proper certificate validation. 

This interaction suggests that Copilot can help improve code security when explicitly prompted to perform security review and refinement. To further examine the effect of such prompts, we analyzed chat sessions from 38 participants whose Copilot interactions were still available. For each session, we appended a follow-up prompt to the end of the chat history, asking Copilot to review the code for security issues and improve it accordingly. We then examined the resulting modifications and recommendations.
In designing this follow-up prompt, we considered two alternatives: (i) D6’s original phrasing, and (ii) a prompt we designed to more explicitly emphasize security best practices: ``\textit{\textbf{Please review the code and improve it based on security best practices}}''. We evaluated both prompts on a code sample containing M2, \textit{trusting all certificates}. In both cases, Copilot successfully corrected the misuse; however, our designed prompt produced more comprehensive security guidance. We therefore adopted this prompt in the subsequent analysis.

\begin{figure*}[!t] 
 \centering
 \includegraphics[width=.9\linewidth]{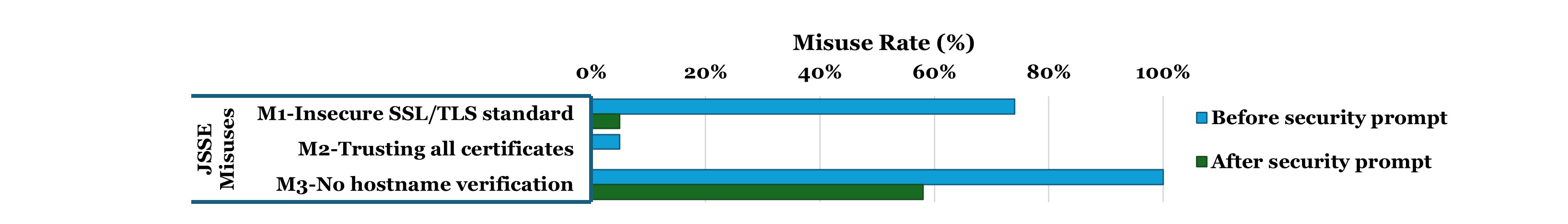}
 \caption{JSSE misuse rates before and after security-oriented prompting.}
 \label{fig:prompt_analysis}
\end{figure*}

The follow-up prompt proved effective in addressing the identified JSSE misuses. For M1, \textit{insecure standard}, Copilot corrected 13 of the 14 instances. In the remaining case, although no fix was applied, Copilot still provided a relevant recommendation: ``\textit{Ensure the application uses a secure version of TLS}''. M2 appeared only once in the AI-assisted condition, and Copilot corrected it successfully. For M3, Copilot corrected 42\% of instances and provided relevant recommendations in 45\% of the remaining cases. For example, in one response it advised, ``\textit{While not implemented here, you should ensure the hostname matches the server's certificate}'', and in another, ``\textit{Ensure the hostname of the server matches the certificate to prevent man-in-the-middle attacks}''. Fig.~\ref{fig:prompt_analysis} compares the rates of JSSE misuses before and after applying the follow-up prompt.

In response to our follow-up prompt for the OAuth task, Copilot consistently recommended securing sensitive data such as secrets and tokens. For instance, it advised: \textit{``Avoid hardcoding sensitive data like client secrets in files''}, addressing M4, and \textit{``If sensitive tokens are stored in the \texttt\small{{DATA\_STORE\_DIR}}, ensure they are encrypted to prevent unauthorized access''}, relevant to M5. However, Copilot could not correct M4 (\textit{local storage of client secrets}), as the secret had already been embedded by the developer, either hardcoded or locally stored. For M5 (\textit{local storage of access tokens}), Copilot suggested hidden files or restrictive file permissions. 
While these practices reflect a degree of security awareness, they remain inadequate and potentially exploitable, especially if an attacker gains elevated privileges (e.g., root access). Copilot did not address or offer recommendations for other OAuth misuses, including \textit{lack of a state parameter}, \textit{lack of SP authentication}, and \textit{lack of PKCE parameters}.

\begin{figure*}
    \centering
    \includegraphics[width=.9\linewidth]{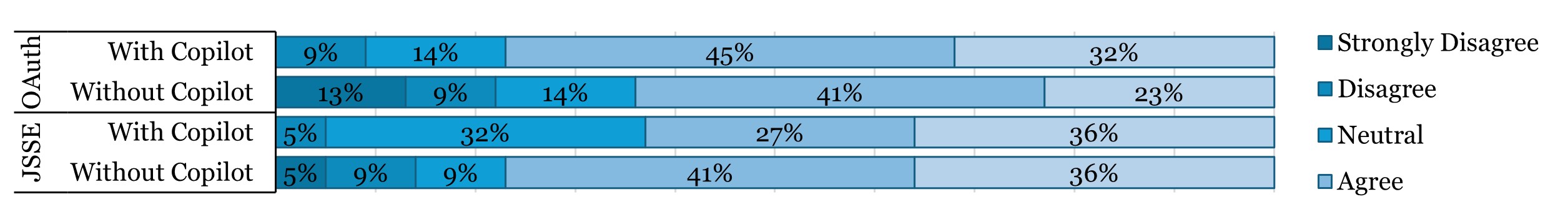}
    \caption{Distribution of participants’ agreement with the statement \textit{“I believe that I solved this task securely”}, across tasks and conditions (5-point Likert scale).}
    \label{fig:self-ass}
\end{figure*}

\paragraph{Self-Assessment Analysis.}
The self-assessments indicate that many participants did not recognize that their implementations remained insecure. Although none of the implementations were fully secure across either task or condition, most participants still expressed confidence in the security of their code. Fig.~\ref{fig:self-ass} shows the distribution of participants' agreement with the statement \textit{``I believe that I solved this task securely''} across tasks and conditions. For the JSSE task, only three participants in the control condition and one participant in the AI-assisted condition explicitly acknowledged that their solution was insecure. Similarly, for the OAuth task, only five participants in the control condition and two participants in the AI-assisted condition identified their code as insecure. These results reveal a clear mismatch between participants’ security self-assessments and the actual security outcomes of their implementations.

This mismatch was observed across both tasks and conditions, suggesting limited awareness of important security requirements even when participants had been informed that their solutions would be evaluated for security. Although fewer participants in the Copilot condition explicitly recognized insecurity, we found no statistically significant effect of Copilot on these self-assessments. When examined by participant background, lower programming experience was associated with lower belief accuracy regarding security, although this trend was not statistically significant.

\begin{findingbox}
\itshape\small
• Participants showed limited security awareness: only two participants explicitly raised security concerns in their prompts, and many did not recognize that their final implementations remained insecure.\\ • Prompting Copilot for security reviews and refinements helped generate fixes or recommendations for certain misuses; however, it lacked awareness of other misuses, particularly in the context of the more complex task.
\end{findingbox}


\section{Discussion} \label{sec:discussion}
This section discusses key findings and implications for developers and researchers.
\subsection{Insights into AI-assisted Security API Use}
Our study shows that Copilot can improve productivity and help developers produce functionally correct code, particularly for more complex security API tasks. This aligns with prior work reporting benefits of AI code assistants for productivity and functional quality~\cite{tabachnyk2022ml, karampatsis2020big, vaithilingam2022expectation, imai2022github, ziegler2022productivity}. However, Copilot’s impact on security was limited. It produced only a slight, statistically non-significant reduction in certain security API misuses.
A key insight from our study is that insecure outcomes were not only a model-output problem, but also a developer-awareness problem. Even though participants were informed that their solutions would be evaluated for security, very few explicitly raised security concerns when engaging with Copilot, and many did not recognize that their final implementations remained insecure. This suggests that persistent security API misuses in AI-assisted development cannot be explained solely by model limitations; they also reflect limited developer security awareness during both code generation and evaluation.
Our findings are broadly consistent with Sandoval et al.~\cite{sandoval2023lost} and Asare et al.~\cite{asare2024user}, who likewise found no significant security improvement from AI assistance. However, our results further suggest that security-oriented prompting can sometimes help Copilot mitigate misuses, especially in the simpler JSSE task. In contrast, Copilot showed limited awareness of several common OAuth misuses even when explicitly prompted, likely reflecting the greater complexity of OAuth flows and the prevalence of insecure patterns in public code.

Compared with Mousavi et al.~\cite{mousavi2024investigation}, who reported a 70\% misuse rate in ChatGPT-generated security API code, our study observed a 100\% misuse rate across the evaluated tasks. While direct comparison is limited by differences in task design, their OAuth task also showed a 100\% misuse rate, consistent with our results. Importantly, the same overall misuse rate was observed in our control condition, suggesting that Copilot did not introduce additional security risk relative to unaided development in our setting. At the same time, secure use of security APIs remained difficult in both conditions, particularly when developers did not engage with the task in a sufficiently security-aware manner.
\subsection{Implications for Developers}
Our study highlights three key takeaways to guide effective 
use of AI assistants:

{\rectnumber{1}} \textbf{\textit{Be explicit about security in prompts.}} Copilot is more likely to generate secure code or offer meaningful security recommendations when prompts contain clear and specific security-related instructions. For example, asking Copilot to ``\textit{review the code and improve it based on security best practices}'' can, in some cases, lead to security improvements or recommendations. Developers should therefore consider explicitly incorporating security considerations into their natural-language interactions with AI tools.

\rectnumber{2} \textbf{\textit{AI assistance is not a substitute for security verification.}} Blind reliance on AI-generated code can introduce serious vulnerabilities into production systems and undermine overall software security. Although Copilot may support certain security best practices when explicitly prompted, it remains far from generating fully secure solutions. Developers must therefore remain vigilant, critically assess AI-generated code, and apply appropriate security measures, especially when handling sensitive data and working with security APIs.


{\rectnumber{3}} \textbf{\textit{Security expertise and continuous learning remain essential.}}
Effective use of AI coding assistants in secure development requires a strong foundation in security principles, reinforced through ongoing security education and hands-on training. Developers must keep pace with evolving standards and best practices, including awareness of deprecated APIs, insecure cryptographic algorithms, and insecure patterns that are no longer recommended, as AI tools may continue to suggest them based on outdated training data.

\subsection{Implications for Researchers}
Our findings highlight several key avenues for advancing research at the intersection of AI-assisted software development and security:


\rectnumber{1} \textbf{\textit{Supporting security-aware engagement with AI tools.}} There is a pressing need to equip developers with the skills and resources required to engage with AI assistants in a security-aware manner when working with security APIs. Many participants in our study neglected security considerations in both their interactions with Copilot and their evaluation of the resulting code, despite being informed that security would be assessed. This highlights the need for practical interventions—such as security training modules, real-time feedback systems, and prompt-design support—that help developers both raise security concerns during interaction and critically assess the security of generated code.

{\rectnumber{2}} \textbf{\textit{Advancing effective misuse detection tools.}} Despite the growing availability of static analysis and program repair tools, existing solutions remain insufficient for detecting and correcting security API misuses~\cite{mousavi2023detecting}. Most tools exhibit limited support for a broad range of security libraries, are prone to high false-positive and false-negative rates, and often provide generic or context-insensitive recommendations. Furthermore, many are restricted to specific programming languages or fail to scale effectively to real-world projects~\cite{zhang2022automatic}. These limitations underscore the need for more accurate and context-aware tools that can support developers in identifying and repairing security API misuses—particularly before integrating AI-generated code into software systems.


{\rectnumber{3}} \textbf{\textit{Improving API usability.}} The default trust manager in JSSE offers an important security advantage by automatically validating certificates against the system trust store. In our study, this secure-by-default feature effectively prevented the common misuse of \textit{trusting all certificates} in both the experimental and control groups. In contrast, hostname verification was frequently misused because it requires explicit developer awareness and manual implementation, which many participants struggled with. This contrast highlights an important implication: \textit{security APIs should be designed with usability as a core principle}. Developers, especially those without deep security expertise, should be able to use these APIs securely. Addressing this challenge requires research to identify usability barriers in current APIs and develop practical, developer-friendly solutions that support secure integration in real-world software systems.

{\rectnumber{4}} \textbf{\textit{Enhancing security of AI-generated code.}} When prompted to review code for security, Copilot could identify and suggest fixes for certain API misuses. However, it failed to detect and address other misuses, particularly for the more complex task in our study. This limitation aligns with the widespread presence of such misuses in real-world code repositories, likely inherited by LLMs trained on these datasets~\cite{mousavi2024investigation}. 
Recent research has investigated a range of techniques to improve LLMs' understanding and handling of software security. These include fine-tuning on security-specific datasets~\cite{elzemity2025cyberllminstructnewdatasetanalysing}, prompt engineering~\cite{bruni2025benchmarking}, incorporating feedback from dynamic testing and static analysis~\cite{dubniczky2025castle}, reinforcement learning with human or automated feedback~\cite{ji2025saferlhfvsafereinforcement}, the integration of security constraints during decoding~\cite{fu2024constrained}, adversarial training~\cite{lu2025adversarialtrainingmultimodallarge}, and modular prefix-based techniques with prompting~\cite{he2023large}.
Future research should extend these efforts to the 
domain of security APIs. In particular, curating high-quality, security-focused datasets that accurately reflect the correct use of security APIs is essential to enable targeted fine-tuning. Moreover, advancing static analysis techniques to more reliably detect and repair security API misuse is a critical step; the feedback they provide can be integrated into the generation process to better guide LLMs toward secure API usage. Additionally, incorporating API specifications, using methods such as Retrieval-Augmented Generation (RAG), offers a promising direction for enriching LLM outputs with relevant security context. Finally, it is imperative to develop mechanisms that ensure LLMs remain aligned with evolving API specifications and emerging security best practices over time.

\vspace{-10pt}
\section{Threats to Validity} \label{sec:threats}
\vspace{-5pt}
This section discusses potential limitations that may affect the reliability and generalizability of our findings, and the steps taken to mitigate them.

\vspace{3pt}
\subheading{Generalizability} 
Our findings may not fully generalize to AI code assistants beyond Copilot or its future evolutions. Nevertheless, Copilot’s widespread adoption among developers makes it a practical and relevant choice for evaluating real-world development workflows. 
Additionally, the study focuses on the Java programming language and two specific security APIs: JSSE and OAuth. These choices were driven by the need to maintain task feasibility within a two-hour session. Although this scope limits applicability to other languages or security APIs, the prevalence of Java (particularly in enterprise and Android development) and the critical role of the selected APIs support the relevance of our findings within common development contexts. Finally, we recruited developers through Freelancer, which may not fully represent the broader developer population. Nevertheless, Freelancer has been recommended by prior research for recruiting participants in security-oriented studies~\cite{kaur2022recruit}, and has also been used in several user studies involving security programming tasks~\cite{naiakshina2020conducting, naiakshina2019if, geierhaas2022let, danilova2020replication}, supporting its suitability for our study.

\vspace{3pt}
\subheading{Study Design Validity}
All participants completed two tasks and experienced the control and experimental conditions, which could introduce learning or fatigue effects. To mitigate this, we counterbalanced both Copilot usage and task order across four groups. Participants in the control condition were instructed not to use any AI-based code assistants; however, we could not guarantee full compliance. To address this, we monitored sessions via screen recording and excluded non-compliant cases from the final analysis. Another potential limitation was varying levels of familiarity with Copilot, which could affect participants’ ability to use it effectively. To reduce this bias, all participants were required to review a brief tutorial on Copilot and its use in VS Code prior to their sessions. They were also allowed to revisit the tutorial during the study if needed. 

\vspace{3pt}
\subheading{Evaluation Validity} Our results might be influenced by the methodology used for misuse detection. Existing automated tools for identifying security API misuses suffer from high false positive and negative rates and lack comprehensive coverage across all security APIs. 
Given these limitations, the first author and a security researcher from our lab, each with four years of security experience, independently reviewed all submissions ($\kappa$ = 0.97), and disagreements were resolved through discussion.
Additionally, our results might be affected by participants’ backgrounds, such as prior familiarity with the specific security APIs used in the tasks. When evaluating functionality, we observed a difference in the OAuth task’s outcomes, where more participants in the experimental condition produced functional code compared to those in the control condition. Our statistical analysis confirmed that this difference remained significant even after controlling for background. Regarding security, only minor variations in a few misuse types were observed across conditions and backgrounds, but these were not large enough to yield meaningful conclusions.


\section{Conclusion} \label{sec:conclusion}
\vspace{-5pt}

This study presents the first empirical investigation into how AI code assistants influence the use of security APIs in real-world development settings. In a within-subject study with 44 professional developers, we compared outcomes on security API programming tasks completed with and without GitHub Copilot assistance. While Copilot improved participants’ ability to produce functionally correct code, particularly for more complex tasks, its support for secure API usage remained limited. No participant produced a fully secure implementation, regardless of AI assistance or background, highlighting the ongoing challenges of secure software development.
Our analysis further shows that security challenges in AI-assisted development are not only a model-output problem, but also a developer-awareness problem. Despite being informed that security would be evaluated, only a few participants explicitly considered security when engaging with Copilot, and many did not recognize that their final implementations remained insecure. Although targeted security-focused prompts could sometimes help Copilot generate more secure revisions or recommendations, Copilot did not reliably identify or mitigate all types of misuses.
These findings underscore the need for both improved tooling and greater developer awareness to enhance the security of AI-assisted development. Future research should explore ways to integrate explicit security guidance into code assistants and design mechanisms that foster a proactive, security-conscious mindset among developers.


\bibliographystyle{unsrt}
\bibliography{references}

@article{mousavi2023detecting,
  title={Detecting misuse of security {APIs}: A systematic review},
  author={Mousavi, Zahra and Islam, Chadni and Babar, Muhammad Ali and Abuadbba, Alsharif and Moore, Kristen},
  journal={ACM Computing Surveys},
  volume={57},
  number={12},
  pages={1--39},
  year={2025},
  publisher={ACM New York, NY}
}

@misc{hardt2012rfc,
    series =    {Request for Comments},
    number =    6749,
    howpublished =  {RFC 6749},
    publisher = {RFC Editor},
    doi =       {10.17487/RFC6749},
    url =       {https://www.rfc-editor.org/info/rfc6749},
    author =    {Dick Hardt},
    title =     {{The OAuth 2.0 Authorization Framework}},
    pagetotal = 76,
    year =      2012,
    month =     oct,
    note         = {Accessed May 26, 2025}
}

@inproceedings{pearce2022asleep,
  title={Asleep at the keyboard? assessing the security of github copilot’s code contributions},
  author={Pearce, Hammond and Ahmad, Baleegh and Tan, Benjamin and Dolan-Gavitt, Brendan and Karri, Ramesh},
  booktitle={2022 IEEE Symposium on Security and Privacy (SP)},
  pages={754--768},
  year={2022},
  organization={IEEE}
}

@article{asare2023github,
  title={{Is GitHub’s Copilot as bad as humans at introducing vulnerabilities in code?}},
  author={Asare, Owura and Nagappan, Meiyappan and Asokan, N},
  journal={Empirical Software Engineering},
  volume={28},
  number={6},
  pages={1--24},
  year={2023},
  publisher={Springer}
}

@inproceedings{khoury2023secure,
  title={{How secure is code generated by chatgpt?}},
  author={Khoury, Rapha{\"e}l and Avila, Anderson R and Brunelle, Jacob and Camara, Baba Mamadou},
  booktitle={2023 IEEE international conference on systems, man, and cybernetics (SMC)},
  pages={2445--2451},
  year={2023},
  organization={IEEE}
}

@inproceedings{siddiq2022empirical,
  title={An Empirical Study of Code Smells in Transformer-based Code Generation Techniques},
  author={Siddiq, Mohammed Latif and Majumder, Shafayat H and Mim, Maisha R and Jajodia, Sourov and Santos, Joanna CS},
  booktitle={2022 IEEE 22nd International Working Conference on Source Code Analysis and Manipulation (SCAM)},
  pages={71--82},
  year={2022},
  organization={IEEE}
}

@article{nair2023generating,
  title={{Generating secure hardware using ChatGPT resistant to CWEs}},
  author={Nair, Madhav and Sadhukhan, Rajat and Mukhopadhyay, Debdeep},
  journal={Cryptology ePrint Archive},
  year={2023}
}

@inproceedings{fahl2012eve,
  title={{Why Eve and Mallory love Android: An analysis of Android SSL (in) security}},
  author={Fahl, Sascha and Harbach, Marian and Muders, Thomas and Baumg{\"a}rtner, Lars and Freisleben, Bernd and Smith, Matthew},
  booktitle={Proceedings of the 2012 ACM conference on Computer and communications security},
  pages={50--61},
  year={2012}
}

@techreport{sakimura2015proof,
  title={Proof key for code exchange by {OA}uth public clients},
  author={Sakimura, Nat and Bradley, John and Agarwal, Naveen},
 institution = {Internet Engineering Task Force (IETF)},
  year={2015}
}

@misc{barnes2015deprecating,
  author={Barnes, Richard and Thomson, Martin and Pironti, Alfredo and Langley, Adam},
  title={Deprecating secure sockets layer version 3.0},
  url={https://tools.ietf.org/html/rfc7568},  
  year={2015},
}

@misc{k2021deprecating,
  author={K. Moriarty and S. Farrell},
  title={Deprecating {TLSv}1.0 and {TLSv}1.1},
  url={https://tools.ietf.org/html/draft-ietf-tls-oldversions-deprecate-12}, 
  year={2021},
  note = {Accessed June 10, 2023}
}

@techreport{turner2011prohibiting,
  title={Prohibiting secure sockets layer {(SSL)} version 2.0},
  author={Turner, Sean and Polk, Tim},
  year={2011}
}

@article{kruger2019crysl,
  title={Cry{SL}: An extensible approach to validating the correct usage of cryptographic {API}s},
  author={Kr{\"u}ger, Stefan and Sp{\"a}th, Johannes and Ali, Karim and Bodden, Eric and Mezini, Mira},
  journal={IEEE Transactions on Software Engineering},
  volume={47},
  number={11},
  pages={2382--2400},
  year={2019},
  publisher={IEEE}
}

@inproceedings{acar2016you,
  title={You get where you're looking for: The impact of information sources on code security},
  author={Acar, Yasemin and Backes, Michael and Fahl, Sascha and Kim, Doowon and Mazurek, Michelle L and Stransky, Christian},
  booktitle={2016 IEEE Symposium on Security and Privacy (SP)},
  pages={289--305},
  year={2016},
  organization={IEEE}
}

@article{zhang2022automatic,
  title={Automatic Detection of {J}ava Cryptographic {API} Misuses: Are We There Yet?},
  author={Zhang, Ying and Kabir, Md Mahir Asef and Xiao, Ya and Yao, Danfeng and Meng, Na},
  journal={IEEE Transactions on Software Engineering},
  volume={49},
  number={1},
  pages={288--303},
  year={2022},
  publisher={IEEE}
}

@inproceedings{rahaman2019cryptoguard,
  title={Cryptoguard: High precision detection of cryptographic vulnerabilities in massive-sized java projects},
  author={Rahaman, Sazzadur and Xiao, Ya and Afrose, Sharmin and Shaon, Fahad and Tian, Ke and Frantz, Miles and Kantarcioglu, Murat and Yao, Danfeng},
  booktitle={Proceedings of the 2019 ACM SIGSAC Conference on Computer and Communications Security},
  pages={2455--2472},
  year={2019}
}

@inproceedings{shernan2015more,
  title={More guidelines than rules: {CSRF} vulnerabilities from noncompliant {OA}uth 2.0 implementations},
  author={Shernan, Ethan and Carter, Henry and Tian, Dave and Traynor, Patrick and Butler, Kevin},
  booktitle={Detection of Intrusions and Malware, and Vulnerability Assessment: 12th International Conference, DIMVA 2015, Milan, Italy, July 9-10, 2015, Proceedings 12},
  pages={239--260},
  year={2015},
  organization={Springer}
}

@inproceedings{egele2013empirical,
  title={An empirical study of cryptographic misuse in android applications},
  author={Egele, Manuel and Brumley, David and Fratantonio, Yanick and Kruegel, Christopher},
  booktitle={Proceedings of the 2013 ACM SIGSAC conference on Computer \& communications security},
  pages={73--84},
  year={2013}
}

@inproceedings{acar2017developers,
  title={Developers need support, too: A survey of security advice for software developers},
  author={Acar, Yasemin and Stransky, Christian and Wermke, Dominik and Weir, Charles and Mazurek, Michelle L and Fahl, Sascha},
  booktitle={2017 IEEE Cybersecurity Development (SecDev)},
  pages={22--26},
  year={2017},
  organization={IEEE}
}

@inproceedings{bianchi2018broken,
  title={{Broken Fingers: On the Usage of the Fingerprint API in Android}},
  author={Bianchi, Antonio and Fratantonio, Yanick and Machiry, Aravind and Kruegel, Christopher and Vigna, Giovanni and Chung, Simon Pak Ho and Lee, Wenke},
  booktitle={NDSS},
  year={2018}
}

@inproceedings{georgiev2012most,
  title={The most dangerous code in the world: validating SSL certificates in non-browser software},
  author={Georgiev, Martin and Iyengar, Subodh and Jana, Suman and Anubhai, Rishita and Boneh, Dan and Shmatikov, Vitaly},
  booktitle={Proceedings of the 2012 ACM conference on Computer and communications security},
  pages={38--49},
  year={2012}
}

@inproceedings{karampatsis2020big,
  title={Big code!= big vocabulary: Open-vocabulary models for source code},
  author={Karampatsis, Rafael-Michael and Babii, Hlib and Robbes, Romain and Sutton, Charles and Janes, Andrea},
  booktitle={Proceedings of the ACM/IEEE 42nd International Conference on Software Engineering},
  pages={1073--1085},
  year={2020}
}

@inproceedings{geierhaas2022let,
  title={$\{$Let’s$\}$ Hash: Helping Developers with Password Security},
  author={Geierhaas, Lisa and Ortloff, Anna-Marie and Smith, Matthew and Naiakshina, Alena},
  booktitle={Eighteenth Symposium on Usable Privacy and Security (SOUPS 2022)},
  pages={503--522},
  year={2022}
}

@inproceedings{al2019oauthlint,
  title={Oauthlint: An empirical study on oauth bugs in android applications},
  author={Al Rahat, Tamjid and Feng, Yu and Tian, Yuan},
  booktitle={2019 34th IEEE/ACM International Conference on Automated Software Engineering (ASE)},
  pages={293--304},
  year={2019},
  organization={IEEE}
}

@inproceedings{naiakshina2020conducting,
  title={On conducting security developer studies with cs students: Examining a password-storage study with cs students, freelancers, and company developers},
  author={Naiakshina, Alena and Danilova, Anastasia and Gerlitz, Eva and Smith, Matthew},
  booktitle={Proceedings of the 2020 CHI Conference on Human Factors in Computing Systems},
  pages={1--13},
  year={2020}
}

@misc{PYPL,
 author = {Pierre Carbonnelle},
 title =  {PYPL Popularity of Programming Language},
 url =    {https://pypl.github.io/PYPL.html},
 year = {2023}, 
 note = {Accessed May 30, 2025}
}

@inproceedings{chen2014oauth,
  title={Oauth demystified for mobile application developers},
  author={Chen, Eric Y and Pei, Yutong and Chen, Shuo and Tian, Yuan and Kotcher, Robert and Tague, Patrick},
  booktitle={Proceedings of the 2014 ACM SIGSAC conference on computer and communications security},
  pages={892--903},
  year={2014}
}

@inproceedings{wang2015vulnerability,
  title={Vulnerability assessment of oauth implementations in android applications},
  author={Wang, Hui and Zhang, Yuanyuan and Li, Juanru and Liu, Hui and Yang, Wenbo and Li, Bodong and Gu, Dawu},
  booktitle={Proceedings of the 31st annual computer security applications conference},
  pages={61--70},
  year={2015}
}

@misc{java_popularity,
  author       = {Melanson, Mike},
  title        = {{Don’t Call It a Comeback: Why Java Is Still Champ}},
  year         = {2022},
  url          = {https://github.com/readme/featured/java-programming-language},
  note         = {Accessed May 30, 2025}
}

@article{sharif2022best,
  title={Best current practices for OAuth/OIDC Native Apps: A study of their adoption in popular providers and top-ranked Android clients},
  author={Sharif, Amir and Carbone, Roberto and Sciarretta, Giada and Ranise, Silvio},
  journal={Journal of Information Security and Applications},
  volume={65},
  pages={103097},
  year={2022},
  publisher={Elsevier}
}

@inproceedings{sandoval2023lost,
  title={Lost at c: A user study on the security implications of large language model code assistants},
  author={Sandoval, Gustavo and Pearce, Hammond and Nys, Teo and Karri, Ramesh and Garg, Siddharth and Dolan-Gavitt, Brendan},
  booktitle={32nd USENIX Security Symposium (USENIX Security 23)},
  pages={2205--2222},
  year={2023}
}

@inproceedings{perry2023users,
  title={{Do users write more insecure code with AI assistants?}},
  author={Perry, Neil and Srivastava, Megha and Kumar, Deepak and Boneh, Dan},
  booktitle={Proceedings of the 2023 ACM SIGSAC Conference on Computer and Communications Security},
  pages={2785--2799},
  year={2023}
}

@article{tabachnyk2022ml,
  title={Ml-enhanced code completion improves developer productivity},
  author={Tabachnyk, Maxim and Nikolov, Stoyan and others},
  journal={Google Research Blog. July},
  volume={26},
  year={2022}
}

@inproceedings{kaur2022recruit,
  title={Where to recruit for security development studies: Comparing six software developer samples},
  author={Kaur, Harjot and Amft, Sabrina and Votipka, Daniel and Acar, Yasemin and Fahl, Sascha},
  booktitle={31st USENIX Security Symposium (USENIX Security 22)},
  pages={4041--4058},
  year={2022}
}

@inproceedings{danilova2020replication,
  title={Replication: On the Ecological Validity of Online Security Developer Studies: Exploring Deception in a $\{$Password-Storage$\}$ Study with Freelancers},
  author={Danilova, Anastasia and Naiakshina, Alena and Deuter, Johanna and Smith, Matthew},
  booktitle={Sixteenth Symposium on Usable Privacy and Security (SOUPS 2020)},
  pages={165--183},
  year={2020}
}

@inproceedings{naiakshina2019if,
  title={"If you want, I can store the encrypted password" A Password-Storage Field Study with Freelance Developers},
  author={Naiakshina, Alena and Danilova, Anastasia and Gerlitz, Eva and Von Zezschwitz, Emanuel and Smith, Matthew},
  booktitle={Proceedings of the 2019 CHI Conference on Human Factors in Computing Systems},
  pages={1--12},
  year={2019}
}

@article{liu2024no,
  title={No need to lift a finger anymore? Assessing the quality of code generation by ChatGPT},
  author={Liu, Zhijie and Tang, Yutian and Luo, Xiapu and Zhou, Yuming and Zhang, Liang Feng},
  journal={IEEE Transactions on Software Engineering},
  year={2024},
  publisher={IEEE}
}

@article{fu2025security,
  title={{Security Weaknesses of Copilot-Generated Code in GitHub Projects: An Empirical Study}},
  author={Fu, Yujia and Liang, Peng and Li, Zengyang and Shahin, Mojtaba and Yu, Jiaxin and Chen, Jinfo},
  journal={ACM Transactions on Software Engineering and Methodology},
  year={2025},
  publisher={ACM New York, NY}
}

@inproceedings{asare2024user,
  title={A user-centered security evaluation of Copilot},
  author={Asare, Owura and Nagappan, Meiyappan and Asokan, N},
  booktitle={Proceedings of the IEEE/ACM 46th International Conference on Software Engineering},
  pages={1--11},
  year={2024}
}

@inproceedings{mousavi2024investigation,
  title={{An investigation into misuse of Java security APIs by large language models}},
  author={Mousavi, Zahra and Islam, Chadni and Moore, Kristen and Abuadbba, Alsharif and Babar, M Ali},
  booktitle={Proceedings of the 19th ACM Asia Conference on Computer and Communications Security},
  pages={1299--1315},
  year={2024}
}

@inproceedings{vaithilingam2022expectation,
  title={Expectation vs. experience: Evaluating the usability of code generation tools powered by large language models},
  author={Vaithilingam, Priyan and Zhang, Tianyi and Glassman, Elena L},
  booktitle={Chi conference on human factors in computing systems extended abstracts},
  pages={1--7},
  year={2022}
}

@inproceedings{imai2022github,
  title={Is GitHub Copilot a substitute for human pair-programming? an empirical study},
  author={Imai, Saki},
  booktitle={Proceedings of the ACM/IEEE 44th International Conference on Software Engineering: Companion Proceedings},
  pages={319--321},
  year={2022}
}

@inproceedings{ziegler2022productivity,
  title={Productivity assessment of neural code completion},
  author={Ziegler, Albert and Kalliamvakou, Eirini and Li, X Alice and Rice, Andrew and Rifkin, Devon and Simister, Shawn and Sittampalam, Ganesh and Aftandilian, Edward},
  booktitle={Proceedings of the 6th ACM SIGPLAN International Symposium on Machine Programming},
  pages={21--29},
  year={2022}
}

@misc{Softonic_2024,
  author = {{Softonic}},
  title = {Microsoft GitHub Copilot: Statistics and Trends},
  year = {2024},
  month = aug,
  url = {https://en.softonic.com/articles/microsoft-github-copilot-statistics-trends},  
  note         = {Accessed May 3, 2025}
}

@article{arani2024systematic,
  title={Systematic literature review on application of learning-based approaches in continuous integration},
  author={Arani, Ali Kazemi and Le, Triet Huynh Minh and Zahedi, Mansooreh and Babar, M Ali},
  journal={IEEE Access},
  volume={12},
  pages={135419--135450},
  year={2024},
  publisher={IEEE}
}

@inproceedings{arani2023sok,
  title={SoK: Machine learning for continuous integration},
  author={Arani, Ali Kazemi and Zahedi, Mansooreh and Le, Triet Huynh Minh and Babar, Muhammad Ali},
  booktitle={2023 IEEE/ACM International Workshop on Cloud Intelligence \& AIOps (AIOps)},
  pages={8--13},
  year={2023},
  organization={IEEE}
}

@misc{hipolito2025github,
  author       = {Melvin Hipolito},
  title        = {GitHub Copilot users surpass 20 million as AI tools surge in demand},
  url   = {https://itbrief.news/story/github-copilot-users-surpass-20-million-as-ai-tools-surge-in-demand},
  year         = {2025},
  month        = {Jul},
  note         = {Accessed January 29, 2026},
}

@misc{ciodive_2024,
  author={CIODIVE},
  title={GitHub Copilot drives revenue growth amid subscriber base expansion},
  howpublished = {\url{https://www.ciodive.com/news/github-copilot-subscriber-count-revenue-growth/706201/}}, 
  year={2024},
  note = {Accessed May 3, 2025}
}

@inproceedings{kholoosi2024qualitative,
  title={A Qualitative Study on Using ChatGPT for Software Security: Perception vs. Practicality},
  author={Kholoosi, M Mehdi and Babar, M Ali and Croft, Roland},
  booktitle={2024 IEEE 6th International Conference on Trust, Privacy and Security in Intelligent Systems, and Applications (TPS-ISA)},
  pages={107--117},
  year={2024},
  organization={IEEE}
}

@misc{elzemity2025cyberllminstructnewdatasetanalysing,
      title={{CyberLLMInstruct: A New Dataset for Analysing Safety of Fine-Tuned LLMs Using Cyber Security Data}}, 
      author={Adel ElZemity and Budi Arief and Shujun Li},
      year={2025},
      eprint={2503.09334},
      archivePrefix={arXiv},
      primaryClass={cs.CR},
      url={https://arxiv.org/abs/2503.09334}, 
}

@article{dubniczky2025castle,
  title={Castle: Benchmarking dataset for static code analyzers and llms towards cwe detection},
  author={Dubniczky, Richard A and Horv{\'a}t, Krisztofer Zoltan and Bisztray, Tam{\'a}s and Ferrag, Mohamed Amine and Cordeiro, Lucas C and Tihanyi, Norbert},
  journal={arXiv preprint arXiv:2503.09433},
  year={2025}
}

@misc{ji2025saferlhfvsafereinforcement,
      title={{Safe RLHF-V: Safe Reinforcement Learning from Human Feedback in Multimodal Large Language Models}}, 
      author={Jiaming Ji and Xinyu Chen and Rui Pan and Han Zhu and Conghui Zhang and Jiahao Li and Donghai Hong and Boyuan Chen and Jiayi Zhou and Kaile Wang and Juntao Dai and Chi-Min Chan and Sirui Han and Yike Guo and Yaodong Yang},
      year={2025},
      eprint={2503.17682},
      archivePrefix={arXiv},
      primaryClass={cs.LG},
      url={https://arxiv.org/abs/2503.17682}, 
}

@misc{lu2025adversarialtrainingmultimodallarge,
      title={Adversarial Training for Multimodal Large Language Models against Jailbreak Attacks}, 
      author={Liming Lu and Shuchao Pang and Siyuan Liang and Haotian Zhu and Xiyu Zeng and Aishan Liu and Yunhuai Liu and Yongbin Zhou},
      year={2025},
      eprint={2503.04833},
      archivePrefix={arXiv},
      primaryClass={cs.CV},
      url={https://arxiv.org/abs/2503.04833}, 
}

@inproceedings{he2023large,
  title={Large language models for code: Security hardening and adversarial testing},
  author={He, Jingxuan and Vechev, Martin},
  booktitle={Proceedings of the 2023 ACM SIGSAC Conference on Computer and Communications Security},
  pages={1865--1879},
  year={2023}
}

@article{bruni2025benchmarking,
  title={Benchmarking Prompt Engineering Techniques for Secure Code Generation with GPT Models},
  author={Bruni, Marc and Gabrielli, Fabio and Ghafari, Mohammad and Kropp, Martin},
  journal={arXiv preprint arXiv:2502.06039},
  year={2025}
}

@article{fu2024constrained,
  title={Constrained decoding for secure code generation},
  author={Fu, Yanjun and Baker, Ethan and Ding, Yu and Chen, Yizheng},
  journal={arXiv preprint arXiv:2405.00218},
  year={2024}
}

@article{sajadi2025llms,
  title={{Do LLMs consider security? an empirical study on responses to programming questions}},
  author={Sajadi, Amirali and Le, Binh and Nguyen, Anh and Damevski, Kostadin and Chatterjee, Preetha},
  journal={Empirical Software Engineering},
  volume={30},
  number={3},
  pages={101},
  year={2025},
  publisher={Springer}
}

@book{lazar2017research,
  title={Research methods in human-computer interaction},
  author={Lazar, Jonathan and Feng, Jinjuan Heidi and Hochheiser, Harry},
  year={2017},
  publisher={Morgan Kaufmann}
}

@misc{material,
  title        = {User study material},
  author = {anonym.},
  howpublished = {\url{https://figshare.com/s/85124bf8ee443ff34dc3}}
}

@book{hosmer2013applied,
  title={Applied logistic regression},
  author={Hosmer Jr, David W and Lemeshow, Stanley and Sturdivant, Rodney X},
  year={2013},
  publisher={John Wiley \& Sons}
}

@inproceedings{klemmer2024using,
  title={Using ai assistants in software development: A qualitative study on security practices and concerns},
  author={Klemmer, Jan H and Horstmann, Stefan Albert and Patnaik, Nikhil and Ludden, Cordelia and Burton Jr, Cordell and Powers, Carson and Massacci, Fabio and Rahman, Akond and Votipka, Daniel and Lipford, Heather Richter and others},
  booktitle={Proceedings of the 2024 on ACM SIGSAC Conference on Computer and Communications Security},
  pages={2726--2740},
  year={2024}
}

@inproceedings{nadi2016jumping,
  title={Jumping through hoops: Why do Java developers struggle with cryptography APIs?},
  author={Nadi, Sarah and Kr{\"u}ger, Stefan and Mezini, Mira and Bodden, Eric},
  booktitle={Proceedings of the 38th International Conference on Software Engineering},
  pages={935--946},
  year={2016}
}

@String{Computing = "Computing" }

@String{Computer = "{IEEE} Computer" }

@String{Springer = "Springer-Verlag" }

\end{document}